\documentclass[12pt]{article}
\usepackage{graphicx}
\usepackage{ifpdf}
\ifpdf  \fi
\usepackage{amsfonts,amssymb}
\usepackage{amsthm}
\usepackage{amsmath}
\usepackage[dvipsnames]{xcolor}
\usepackage[ruled,vlined]{algorithm2e} 
\usepackage[letterpaper, total={6.5in, 9in}]{geometry}
\usepackage{lineno}
\modulolinenumbers[5] 

\everymath{\displaystyle} 
\usepackage{adjustbox}
\usepackage{soul}

\usepackage{subcaption}
\usepackage[colorinlistoftodos]{todonotes}

\usepackage{mathptmx,helvet,courier,makeidx,multicol,footmisc}
\usepackage[numbers]{natbib}
\bibpunct{(}{)}{;}{a}{,}{,}
\usepackage[bookmarks=false, pdfauthor={Irene Martinez}]{hyperref}
    \hypersetup{colorlinks,
      linkcolor=blue,
      citecolor=Emerald,
      urlcolor=blue}

\usepackage[colorinlistoftodos]{todonotes}

\usepackage{nomencl}
\makenomenclature

\newcommand{\commentout}[1]{}

\newcommand{\ba}{\begin{array}}
        \newcommand{\ea}{\end{array}}
\newcommand{\bc}{\begin{center}}
        \newcommand{\ec}{\end{center}}
\newcommand{\bdm}{\begin{displaymath}}
        \newcommand{\edm}{\end{displaymath}}
\newcommand{\bds} {\begin{description}}
        \newcommand{\eds} {\end{description}}
\newcommand{\ben}{\begin{enumerate}}
        \newcommand{\een}{\end{enumerate}}
\newcommand{\beq}{\begin{equation}}
        \newcommand{\eeq}{\end{equation}}
\newcommand{\bfg} {\begin{figure}}
        \newcommand{\efg} {\end{figure}}
\newcommand{\bi} {\begin {itemize}}
        \newcommand{\ei} {\end {itemize}}
\newcommand{\bqn}{\begin{eqnarray}}
        \newcommand{\eqn}{\end{eqnarray}}
\newcommand{\bqs}{\begin{eqnarray*}}
        \newcommand{\eqs}{\end{eqnarray*}}
\newcommand{\bsl} {\begin{slide}[8.8in,6.7in]}
        \newcommand{\esl} {\end{slide}}
\newcommand{\bsq}{\begin{subequations}}
        \newcommand{\esq}{\end{subequations}}       
\newcommand{\bss} {\begin{slide*}[9.3in,6.7in]}
        \newcommand{\ess} {\end{slide*}}
\newcommand{\btb} {\begin {table}}
        \newcommand{\etb} {\end {table}}

\newcommand{\m}{\mbox}

\newcommand{\reff}[1] {{{\textbf{Figure} \ref {#1}}}}
\newcommand{\refe}[1] {{(Eq. \ref {#1})}}
\newcommand{\reft}[1] {{{\textbf{Table} \ref {#1}}}}

\def\log        {{\m{log}}}
\def\pmb#1{\setbox0=\hbox{$#1$}%
   \kern-.025em\copy0\kern-\wd0
   \kern.05em\copy0\kern-\wd0
   \kern-.025em\raise.0433em\box0 }

\newtheorem{theorem}{Theorem}
\newtheorem{corollary}{Corollary}[theorem]
\newtheorem{lemma}[theorem]{Lemma}

\usepackage[draft]{changes}
\definechangesauthor[name={IM},color=red]{IM}

\usepackage[colorinlistoftodos]{todonotes}

\graphicspath{ {figures/} }

\usepackage{multirow}
\usepackage{float}

\usepackage{setspace}   

\setcitestyle{authoryear, open={((},close={)}}

\title{Priority Queue Formulation of Agent-Based Bathtub Model for Network Trip Flows in the Relative Space}
\author{Irene Mart\'inez\footnote{Corresponding author. TU Delft, Stevinweg 1, Delft 2628 CN, The Netherlands E-mail address: I.MartinezJosemaria@tudelft.nl.} and Wen-Long Jin}
\begin{document}
\maketitle


\begin{abstract}
	\noindent Agent-based models have been extensively used to simulate the behavior of travelers in transportation systems because they allow for realistic and versatile modeling of interactions. However, traditional agent-based models suffer from high computational costs and rely on tracking physical locations, raising privacy concerns. This paper proposes an efficient formulation for the agent-based bathtub model (AB$^2$M) in the relative space, where each agent's trajectory is represented by a time series of the remaining distance to its destination. The AB$^2$M can be understood as a microscopic model that tracks individual trips' initiation, progression, and completion and is an exact numerical solution of the bathtub model for generic (time-dependent) trip distance distributions. 
 The model can be solved for a deterministic set of trips with a given demand pattern (defined by the start time of each trip and its distance), or it can be used to run Monte Carlo simulations to capture the average behavior and variation stochastic demand patterns, described by probabilistic distributions of trip distances and departure times. 
 To enhance the computational efficiency, we introduce a priority queue formulation for AB$^2$M, eliminating the need to update trip positions at each time step and allowing us to run large-scale scenarios with millions of individual trips in seconds. We systematically explore the scaling properties of AB$^2$M and discuss the introduction of biases and numerical errors. Finally, we analyze the upper bound of the computational complexity of the AB$^2$M and the benefits of the priority queue formulation and downscaling on the computational cost.
 The systematic exploration of scaling properties of the modeling of individual agents in the relative space with the AB$^2$M  further enhances its applicability to large-scale transportation systems and opens up opportunities for studying travel time reliability, scheduling, and mode choices. 
\end{abstract}


\section{Introduction}\label{sec:intro}

Transportation systems are characterized by a set of complex interactions between the supply of the system and the demand for goods and travelers. The heterogeneous behavior of travelers as they navigate their dynamic environment leads to complex collective systems' dynamics.
For this reason, agent-based modeling tools, initially developed in the field of computing, are an appropriate method to simulate the complex behavior of travelers \citep{crooks2018agent}. 
The idea behind agent-based modeling is to break down complex systems into individual agents driven by certain rules, such as activity constraints and travel preferences \citep{Bonabeau2002Agent}. Each agent is a distinct, autonomous organism with unique aims and behavior that can adapt its response to changing circumstances.
As a result, agent-based models (ABM) can represent a variety of interactions between various entities in a realistic manner (such as interactions between vehicles and passengers). The versatility of this modeling approach also makes it simple to alter the model assumptions and requirements.
In the transportation domain, ABMs have been traditionally used for microscopic traffic and macroparticles simulations, e.g., with MATSim \citep{balmer2008agent} or POLARIS \citep{AULD2016101}, and have been used to study the system dynamics and to evaluate the impact of management strategies.
The readers are referred to a recent review by  \cite{bastarianto2023agentbased}. However, ABMs also come with some challenges, including data collection and accuracy, very high computational cost \citep{KAGHO2020Agent}.
To reduce the computational cost of simulations, traditional ABM relies on downscaling methodologies \citep{Nicolai2012Using, BENDOR2021Population}, which have been extensively used in the literature.  
For example, the downscaling approach in MATSim software relies on a method that scales the network characteristics based on the reduced number of agents modeled. Although this methodology has been used for years, there is limited systematic understanding as to whether a downsized model can reproduce the same exact results as a full-scale model.
Recent studies show that downscaling ``too much'' might introduce bias into the results \citep{LLORCA2019Effects, BENDOR2021Population}.
Furthermore, even relying on downscaling, the computational complexity of traditional ABM is still high; for example, a MATSim simulation for the city of Paris (with 10\% of the population) takes 5 hours to run on a modern cluster \citep{Horl2019Dynamic}.

Traditionally, ABM models (as well as other traditional traffic flow simulation software) require a generation of a network with individual links that compose the physical configuration. We refer to this traditional perspective as the absolute space.  This approach involves significant computational efforts, both for calibration and simulation. In particular, using ABM to track the movement of people in the absolute space leads to personally identifiable location information being traced, which can lead to privacy concerns. 
In contrast, one can take the relative space approach, where the network is an undifferentiated unit, and disregard the physical locations of vehicle-trips inside the network. In this relative space dimension, the trip flow dynamics are modeled based on their remaining trip distances to their respective destinations, preserving personally identifiable location information.

The bathtub model (a.k.a. reservoir model) is an aggregated model  used to describe the trip flow dynamics in the relative space and has been garnering interest among the transportation research community \citep{Vickrey2020congestion,smallChu2003,Daganzo2007,jin2020generalized,Johari2021}. 
It can be viewed as a network queuing system, and 
it relies on three main assumptions: 
(A1) the network is treated as an undifferentiated unit, where links and individual trips' origins, destinations, and routes are implicit; (A2) the demand is described by the trip distance and the trip initiation rate; and (A3) there exists a network-level speed-density relation, which is nowadays commonly referred to as network fundamental diagram (NFD) or (speed) macroscopic fundamental diagram (MFD). This network-wide relation between speed and density was proposed and calibrated by \cite{Godfrey1969}. Later, the NFD was also studied by \cite{Mahmassani1987} and \cite{Daganzo2007} and has thereafter gained significant interest in the research community. The existence of such relation at the network level through a large scale loop detector data collection  was first verified in downtown Yokohama   by \cite{GEROLIMINISandDaganzo2008}. Thereafter, multiple NFDs have been experimentally verified in several cities \citep{Johari2021}. 
Note that bathtub models can also be viewed as compartmental models \citep{Jin2021Compartmental}, where all the trips in the same network are lumped together in a bathtub.

Most of the bathtub models in the literature assume continuum travelers, like water flowing in and out of a bathtub, and we will refer to them as \textit{continuum} bathtub models.
Thus, it can be viewed as a network queue within the relative space dimension and can be used to model different modes of transportation, e.g., privately operated vehicles (POVs) on the road, shared mobility systems, pedestrians, and other active modes. 
According to \cite{Macal2010Agent} ``every well-formulated system dynamics model has an equivalent formulation as [... a] time-stepped agent-based simulation model [...]".  Further, he shows how some ABM are equivalent to the system dynamics model without providing additional information or advantages, 
and other formulations may  provide more information than a continuum system dynamics model.
In the last decade, several authors have suggested the idea of using agents to model the trip flow dynamics at the network level \citep{ARNOTT2013, DAGANZOandLEHE2015, MARIOTTE2017, Lamotte2018}. \cite{ARNOTT2013} originally proposed in a footnote to use a model in which each distance traveled by commuters should coincide with the distance traveled during their time in the network, but did not propose a method to solve the model, since it was identified as a delay-differential equation.  
Later, the so-called trip-based model (TBM)  was proposed by \cite{MARIOTTE2017} to numerically capture the progression of individual trips. However, existing formulations do not take advantage of the characteristics of the relative space for an efficient algorithm, and many consider some limiting assumptions about the demand, e.g., time-independent trip distances and/or deterministic demand. 
Therefore, there is a need to develop an efficient model in the relative space to capture the individual agents' trajectories for any demand assumption accurately and efficiently. 

In this paper, we propose a highly efficient (simulation) model to capture the agent's dynamics in the relative space. 
We refer to the proposed model as \textit{agent-based} bathtub model (AB$^2$M) for two reasons. First, to highlight that the travelers are people with decision capabilities and can be modeled as independent agents with corresponding socio-economic characteristics. Second, to highlight that the model is based on the same three assumptions of relative space as the bathtub model.  
Unlike continuum bathtub models that rely on aggregated variables, the agent-based formulation provides detailed information on individuals, including travel times. This enables the introduction of heterogeneity and studying higher order moments such as travel time variation, which is crucial for assessing travel time reliability and investigating scheduling \citep{Noland2002Travel} and route choices \citep{LAM2001thevalue}.  
The proposed AB$^2$M  is an exact numerical solution that extends the TBM \citep{ARNOTT2013, FOSGERAU2015, MARIOTTE2017} by accommodating agents with varying trip distances and time-dependent trip distance distributions. Moreover, the AB$^2$M can handle stochastic demands through Monte Carlo simulations. 
To enhance the efficiency of AB$^2$M, we propose a priority queue formulation, which eliminates the need to update the position of trips at each time step, resulting in significant computational performance improvements. For instance, simulations involving one million individual trips over a one-hour period can be processed in less than two seconds. Further, we explore its scaling properties systematically, similar to the downscaling strategies used in traditional ABM models.

The rest of the paper is structured as follows. Section \ref{sec:model} reviews existing models in the literature and proposes the AB$^2$M, which tracks the trajectories of agents in the relative space. Then,  
Section \ref{sec:efficiency} presents a new formulation that is computationally more efficient. 
Section \ref{sec:scalability} discusses the scalability property of the bathtub model, and Section \ref{sec:complexity} reviews the numerical complexity of the proposed models. 
Finally, Section \ref{sec:conclusion_ABM1} concludes the paper and discusses future directions.

\section{Model formulation}\label{sec:model}

\subsection{Review of existing models}\label{sec:review}

The network dynamics in an urban region have been modeled using aggregated models that capture the  inflow and outflow of trips in a system (or region). As mentioned in Section \ref{sec:intro}, the existence of a uni-modal, low-scatter speed density relationship between the speed and average density of the region \citep{Godfrey1969,GEROLIMINISandDaganzo2008,Johari2021}, has allowed the development of continuum bathtub models.

\begin{table}\caption{ \small List of notations.\label{tab:ABBM1:symbols}}
\centering
\small
\begin{tabular}{cl}
\hline
Variable & Description  \\ \hline \hline
$\tilde{D} (t)$  & Average trip distance of entering trips  at time $t$      \\
$E (t)$  & Cumulative number of starting trips  at time $t$      \\
$G (t)$ & Cumulative number of finished trips  at time $t$  \\
$I$ & Total number of trips \\ 
$L_N$  & Total network length   \\
$N(t,i)$ & Position (order) of trip $i$ at time $t$ regarding its characteristic  trip distance. \\ 
$T (i)$  & Trip $i$ start time on simulation\\
$\hat{T} (i)$ & Time when trip $i$ exits the network     \\ 
$V(\rho)$  & Network level speed-density relation \\
$X (i)$ & Trip distance of trip $i$     \\ \hline
$e (t)$  & Trip initiation rate  at time $t$      \\
$g (t)$ & Completion rate of trips  at time $t$  \\
$i$ & Trip ID, which refers to the trip position regarding its initiation time\\ 
$m(t)$ & Total remaining distance to be traveled by active trips\\ 
$t_f$ & Total simulation time \\ 
$u_f$ & Free-flow speed \\ 
$v (t)$ & Average travel speed in the system at time $t$    \\ 
$w$ & Shock wave speed    \\ 
$x (t,i)$  & Remaining distance of trip $i$ at time $t$ \\ 
$y (t,n)$  & Remaining distance of $n$-shortest trip at time $t$ (ordered remining trip distance)\\ 
$z (t)$ & Characteristic network traveled distance      \\ \hline 
$\Delta t $ & Time interval or time step of the simulation. Note that it is not necessarily fixed. \\
$\Delta v $ & Speed variation in a time-step considered. \\ 
$\Theta (t,n)$ & Ordered characteristic  trip distance of active trips at time $t$    \\  \hline
$\delta (t)$ & Active number of trips at time $t$      \\ 
$\gamma (i)$ & Travel time of trip $i$.  \\ 
$\rho (t)$ & Per-lane density in the system at time $t$      \\ 
$\rho_j $ & Per-lane jam density      \\ 
$\theta (i)$ & Characteristic  trip distance of trip $i$      \\ 
$\tilde \varphi(t,x)$ & Joint distribution of trip distance, $x$, and departure times, $t$     \\ \hline
\end{tabular}
\end{table}

In the relative space the demand is defined with the joint distribution of trip distance and departure time \citep{jin2020generalized}, which is described by $\tilde \varphi_{T, X}(t,x)$ and is defined for $x\geq 0$ and $t \geq 0$.  For the readers' convenience  \reft{tab:ABBM1:symbols} presents a list of variables.  
This joint distribution can be continuous or discrete; and deterministic or probabilistic, both in time $t$ and trip distance $x$. Therefore, there are eight possible types of demand in the relative space. 
If the demand is stochastic, the joint distribution $\tilde \varphi_{T,X}(t,x)$ becomes a probabilistic function. 
From the axiom of probability, the joint distribution can be written as the probability of departure time multiplied by the conditional distribution of trip distances, i.e.,  
\begin{equation} \label{eq/joint}
   \tilde \varphi_{T,X}(t,x) = \tilde \varphi_T(t) \cdot \tilde \varphi_{X|T}(x|t),
\end{equation}
where $\tilde \varphi_T(t)$ is the marginal probability density (or mass) function \citep{Martinez2021On}. The conditional distribution $\varphi_{X|T}(x|t)$ is the most generic expression for capturing trip distance distributions as a function of time. 
If the trip distance distributions is a continuous variable, the trip initiation rate is defined by
\begin{equation}\label{eq/e}
    e(t) = I \int_0^{\infty} \tilde \varphi_{T,X}(t,x) dx,
\end{equation}
where $I$ is the total number of trips in the study period. The completion rate of trips, $g(t)$, is determined by the interaction of supply and demand. The supply is defined by the fundamental relation of speed and density, $V(\rho)$, at the network level, i.e., the NFD. Several modeling formulations have been proposed in the literature based on the assumptions of the demand and trip distances, which we summarize in \reft{tab:classification_demand_types}, and described in the rest of the section. 

The trip flow dynamics in continuum bathtub models are generally described by the rate of change of active trips $\dot \delta(t)$. The first derivation of a bathtub model was build on a particular case of A2, i.e., that trip distances follow a (time-independent) negative exponential (NE) distribution, and led to an ordinary differential equation (ODE) \citep{Vickrey1991,Vickrey2020congestion} 
\begin{equation}\label{gbm:vbm}
\dot \delta(t) =  e(t) -  \frac{\delta(t)}{\tilde D}V\left(\frac{\delta(t)}{L_N}\right) ,
\end{equation}
where $L_N$ is the total network lane distance, $V\left(\frac{\delta(t)}{L_N}\right)$ is the speed-density relationship defined by the NFD, and $\tilde D$ is the average (time-independent) trip distance. We will refer to this version as Vickrey's bathtub model (VBM). Similar ODEs where derived independently in the early 2000s by other authors \citep{smallChu2003, Daganzo2007}, but the assumption regarding the trip distance distribution was omitted or only indirectly mentioned. In particular, \cite{Daganzo2007} assumed ``the average trip length is the same for all origins, d''. This ODE formulation is the most extended in the literature \citep{Johari2021}, and sometimes is referred to as accumulation-based model \citep{MARIOTTE2017} or PL model \citep{SIRMATEL2021Modeling}. Other researchers proposed the basic bathtub model (BBM), which assumes that all travelers have the same (homogeneous) trip distance \citep{ARNOTT2013, Arnott2016, ARNOTT2018}. \cite{ARNOTT2013} first suggested that a proper model should ensure that
\begin{equation}\label{eq:TBM0}
    X = \int_T^{T+ \gamma} V\left(\frac{\delta(s)}{L_N} \right) ds,
\end{equation}
where $\gamma$ is the travel time of the vehicle that starts its trip of distance $X$ at time $T$. To solve this integral version, they proposed a formulation which leads to a delay-differential equation, which is mathematically challenging to solve. 
Several authors noted that using \refe{gbm:vbm} to model the dynamics of a system with homogeneous trip distances would lead to inconsistencies, such as instantaneous increase in completion rate of trip with an increased inflow to the system \citep{Arnott1993, MARIOTTE2017,Johari2021}.  

\begin{table}
    \caption{Characteristics of demand assumption in the relative space based on model formulations. TDD: Trip distance distribution, C: Continuous, D: Discrete}
    \centering
    \begin{tabular}{l c c c c c c}
         Characteristic                     & BBM & VBM/ PL & GBM & TBM & MM  & AB$^2$M \\ \hline
        Travelers                              & C & C    & C   & D & C  & D \\
         Time-dependent TDD                    & No         & No            & Yes  & (Generally) No & No & Yes\\
         Trip distance                          & D  & C & D or C & D or C & D or C & D or C\\ \hline

    \end{tabular}
    \label{tab:classification_demand_types}
\end{table}

Instead of relying on continuum models, an alternative is to use agents (discrete travelers) to model the trip flow dynamics. 
\cite{DAGANZOandLEHE2015} were the first to do so. They indirectly considered the relative space, claiming that ``since there are no routes in the macroscopic theory [...] a network governed by an MFD [...] may be modeled as a simple, aspatial queuing system [...]''. In other words, they abstracted the network to a queuing system and explicitly disregarded the (traditional) absolute space dimension. They outlined the idea that the trip dynamics can be modeled as a non-FIFO, multi-channel queuing system, where the trip distance is the customers' workloads, and the servers process the customers at a common rate (the speed determined by the NFD). 
They considered the departure time and trip distance, which they assumed follows a uniform distribution, to be an input to the simulation.
Another well-established formulation is the TBM by \cite{MARIOTTE2017}, where they proposed two numerical resolutions. First, a formulation where the trips are also considered a continuum variable and the cumulative curves of trip initiation and trip completion are calculated. To do so, they estimate the time difference between exit time of vehicle $n$ and vehicle $n+\Delta n$. They proposed a resolution of the method at higher orders and concluded that a first-order approximation is accurate enough. Second, an event-based formulation inspired by an unpublished work by \cite{Lamotte2016}, where the progression of each trip towards the destination is updated every event, i.e., when a vehicle enters or exists the system. Note that in their numerical resolutions, \cite{MARIOTTE2017} assumed that all vehicles have the same trip distance, i.e., a discrete version of BBM. They also considered other demand patterns where the population was divided into different groups with different trip distances. The algorithm proposed updates the traveled distance for all the circulating vehicles during each event. Later, \cite{Lamotte2018} proposed a complete agent-based simulation and to limit the computational complexity they did not consider that each agent may have a different trip distance, but consider 1000 deterministic  trip distances associated each to 1000 batches of agents.  \cite{Lamotte2018} also consider multiple trip distance distributions, including uniform and mixtures of uniform distributions by varying the standard deviation, but all their trip distance distributions had the same (time-independent) average trip distance $\tilde D$.
They also presented a modification of the trip completion rate (a.k.a. outflow function) to capture more generic trip distance distributions, as follows
\begin{equation}\label{eq:TBM2}
    \dot \delta (t) = e(t) - V(\frac{\delta(s)}{L_N} )  \int_0^t e(s) \tilde \varphi \left( \int_s^t V(\frac{\delta(\tau)}{L_N} ) d\tau  \right) ds,
\end{equation}
where $\tilde \varphi$ is the probability density function of the trip distances, which they assumed time-independent. 
This formulation has also been referred as TBM \citep{SIRMATEL2021Modeling}. A more general formulation of the TBM was recently proposed by \cite{LAVAL2023Effect}, where $\tilde \varphi$ is explicitly considered time dependent.
Note that all the above formulations disregard the space dimension \citep{Johari2021}.

Later, \cite{jin2020generalized} introduced the (relative) space dimension into the bathtub model explicitly based on \cite{Vickrey2020congestion} and proposed the generalized bathtub model (GBM) to capture the network trip dynamics for generic demand patterns.  \cite{jin2020generalized} introduces a new state variable, the number of active trips at time $t$ with remaining distance not smaller than $x$, $K(t,x)$, and proposed a partial differential equations (PDEs) instead of ODEs to track the dynamics as
\begin{equation}\label{eq:gbm}
    \frac{\partial K(t,x)}{\partial t} - V\left(\frac{K(t,0)}{L_N}\right) \frac{\partial K(t,x)}{\partial x} = e(t) \tilde \Phi (t,x),
\end{equation}
where $K(t,0) = \delta(t)$ and $\tilde \Phi (t,x)= \int_x^\infty \tilde \varphi(t,y) dy$ is the cumulative density function of entering trips with distances not smaller than $x$. The integral version of the GBM is
\begin{equation}\label{eq:gbm:int}
    \delta(t) = K(t,0) = K(0,z(t)) + \int_0^t e(s) \tilde \Phi(s, z(t) - z(s) ) ds,
\end{equation}
where $z(t)= \int_0^t V(\delta(s)/L_N) ds$ is the characteristic travel distance.
Note that \refe{eq:gbm:int} is a generalization for time-independent trip distance distribution and non-empty initial network of the integral version of \refe{eq:TBM2}, i.e.,
\begin{equation}\label{eq:tbm:integral}
    \delta (t) = \int_0^t e(t) \tilde \Phi\left(\int_s^t V(\delta(\tau)/L_N) d \tau \right) ds.
\end{equation}

Recently, \cite{SIRMATEL2021Modeling} proposed the M-model, which was originally presented by \cite{Lamotte2018}. Based on the conservation of trip-distances, i.e.,
\begin{equation}\label{eq-def-m}
    \dot m (t) = e(t) \tilde D (t) - \delta(t) V\left(\frac{\delta(t)}{L_N}\right),
\end{equation}
they propose to use the total remaining distance to be traveled $m(t)$ to approximate the completion rate of trips with the following dynamic model
\begin{equation}\label{eq:MM}
    \dot \delta(t) =  e(t) -  \frac{\delta(t)}{\tilde D}V\left(\frac{\delta(t)}{L_N}\right)  \left( 1 - \alpha (\frac{m(t)}{\delta (t) D^*} - 1) \right),
\end{equation}
where $D^*$ is the average remaining distance to be traveled in steady states and $\alpha$ is a parameter of the model and should be calibrated from empirical data. Note that $\frac{m(t)}{\delta(t)}= D(t)$ is the average remaining distance to be traveled. Based on the assumption of time-independent trip distance distributions, \cite{SIRMATEL2021Modeling} establish a relationship between the average trip distance $\tilde D$ and the average remaining trip distance $D^*$ for steady states. Note that the M-model \refe{eq:MM} during steady states is simplified to VBM, i.e., \refe{gbm:vbm}. The M-model is an approximation of the dynamics outside of steady state conditions. It has the advantage that one is not required to track the distribution of remaining trip distances, but it requires empirical data to estimate $\alpha$.

Some researchers have concluded that the VBM formulation \refe{gbm:vbm} and the TBM \refe{eq:TBM0} \refe{eq:TBM2} are equivalent when the trip distance distributions follows a time-independent NE distribution or under steady states \citep{Lamotte2018,jin2020generalized,LAVAL2023Effect}. However, the definition of steady (or stationary) state is not consistent in the literature. It can be generally interpreted as  vehicle inflow equals vehicle outflow, i.e., $\dot \delta (t) =0$.  \cite{LAVAL2023Effect} defines the steady state as “[...] the circulating flow over the total network distance, has to match the incoming production [of trip distance]”, i.e. $\dot m (t) = 0$, while \cite{jin2020generalized} defines stationary states as $\frac{\partial K(t,x)}{\partial t}$.

It is worth noting that many of the aforementioned models have been used to study multiple transportation research questions. For example, to study the departure time problem different researchers have used the TBM \citep{Lamotte2018_Morningcommute}, the BBM \citep{ARNOTT2018} or the GBM \citep{Ameli2022Departure}. Moreover, the VBM has been extensively used to propose management strategies at the network level, such as gating strategies \citep{Keyvan-Ekbatani2012} or pricing \citep{ZHENG2012, ZHENG2016Modeling}.

\subsection{Naive Formulation of AB$^2$M}\label{sec:formualtion}

Most of the models reviewed in Section \ref{sec:review} consider continuum travelers, and we refer to them as \textit{continuum} bathtub models.
Instead, in this paper, we are interested in proposing a new formulation to efficiently model the agent-based version of the bathtub model, i.e., considering discrete travelers. To the best of our knowledge, only three of the aforementioned studies proposed such a discrete formulation and study the models characteristics \citep{DAGANZOandLEHE2015, MARIOTTE2017,Lamotte2018}. \cite{DAGANZOandLEHE2015} proposed a discrete time step simulation and both \cite{MARIOTTE2017} and \cite{Lamotte2018} proposed event-based simulations. In all three studies the authors track the commuters' traveled distance and the trips are marked as completed when the traveled distance equals or exceeds their trip distance. In this section, we will present a formalized agent-based bathtub algorithm with fixed time step that is conceptually equivalent to existing models in the literature.

First, we introduce the characteristic network traveled distance as 
\begin{equation}\label{eq/cumtravel}
    z(t) = \int_0^t v(s) ds,
\end{equation}
which as defined by \cite{Lamotte2018} as ``cumulative distance traveled'' and by \cite{jin2020generalized} as the ``characteristic travel distance''.  This  characteristic network traveled distance is useful to simplify the notation of the remaining trip distance. 
From the agent-based perspective, one can interpret the characteristic network traveled distance as the cumulative traveled distance by a reference vehicle-trip that initiates its trip at $t=0$ and  never exits the system \citep{Lamotte2018}.

In the AB$^2$M, the active number of agents in the system is not tracked by differential of integral equations as presented in Section \ref{sec:review}.
Instead, the trip progression of each individual agent $i$ (towards their destination) is tracked by reducing her remaining trip distance over time. 
Each vehicle-trip represents an agent moving in the relative space, and the coordinates will be Lagrangian coordinates.  Conceptually, the AB$^2$M is similar to a traditional microscopic simulation model,  but (i) the trajectory is in the relative space and the personally identifiable location information is preserved, and (ii) the speed is determined by a global speed-density relation (NFD) instead of a local speed-density relation.

\begin{figure}
    \centering
    \includegraphics[width=\textwidth]{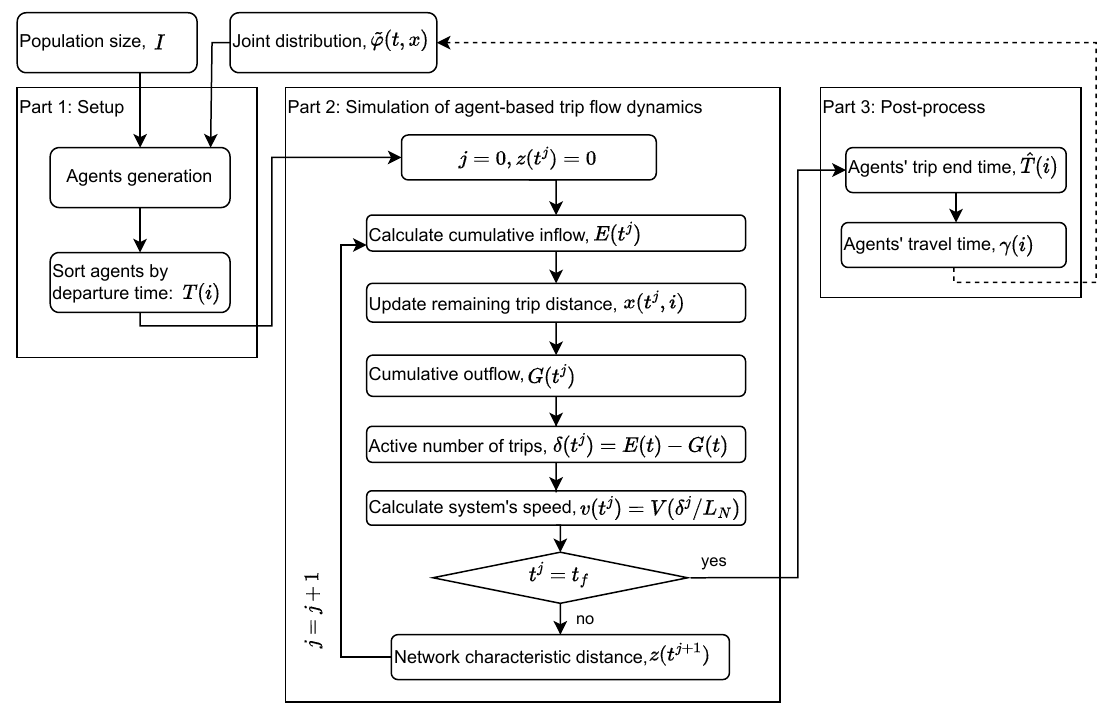}
    \caption{ \small Overview of the main steps of the AB$^2$M.}
    \label{fig:overview}
\end{figure}

\begin{figure}
\centering
\begin{subfigure}{.32\textwidth}
\includegraphics[width=1\textwidth]{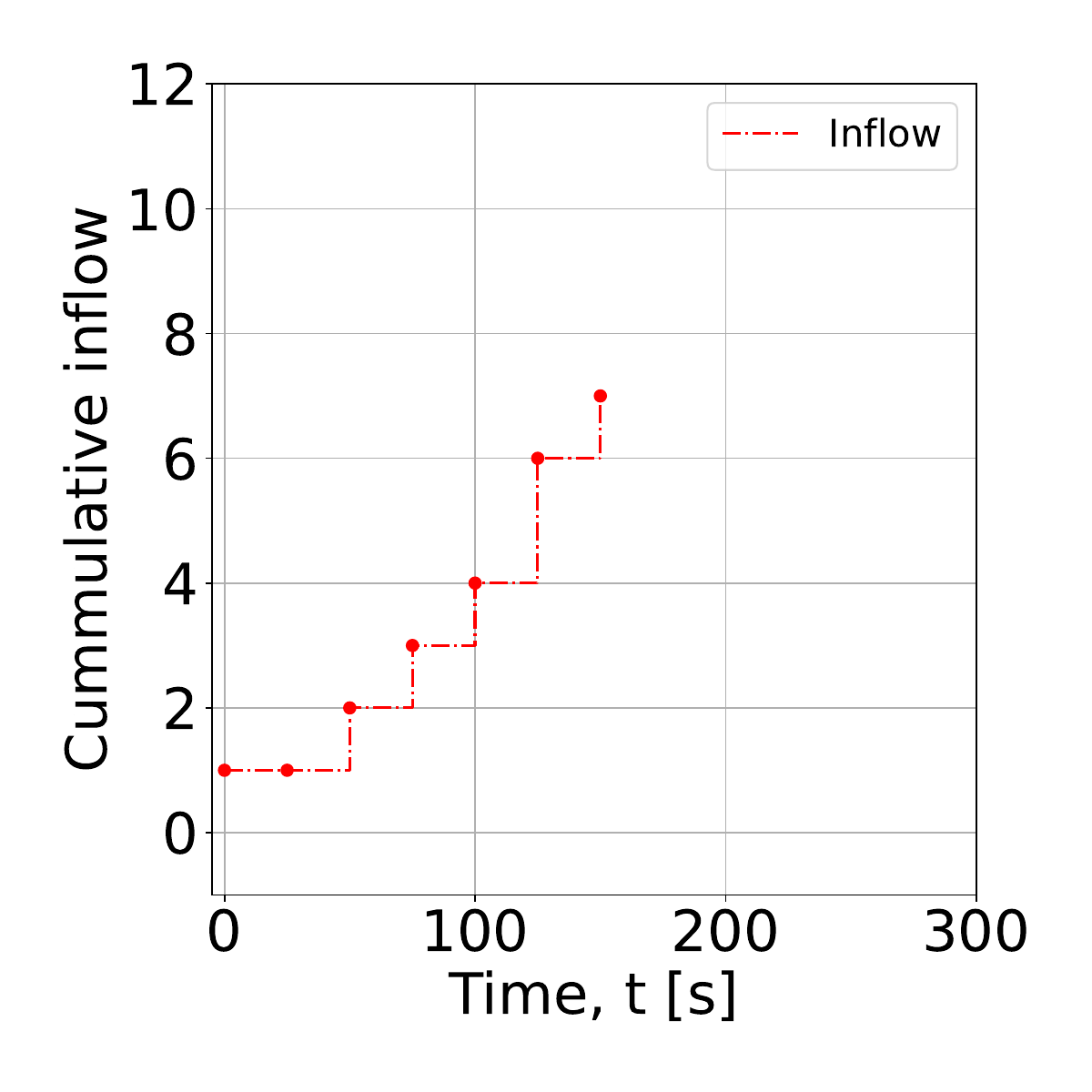}
\caption{ \small } 
\end{subfigure}%
\begin{subfigure}{.32\textwidth}
\includegraphics[width=1\textwidth]{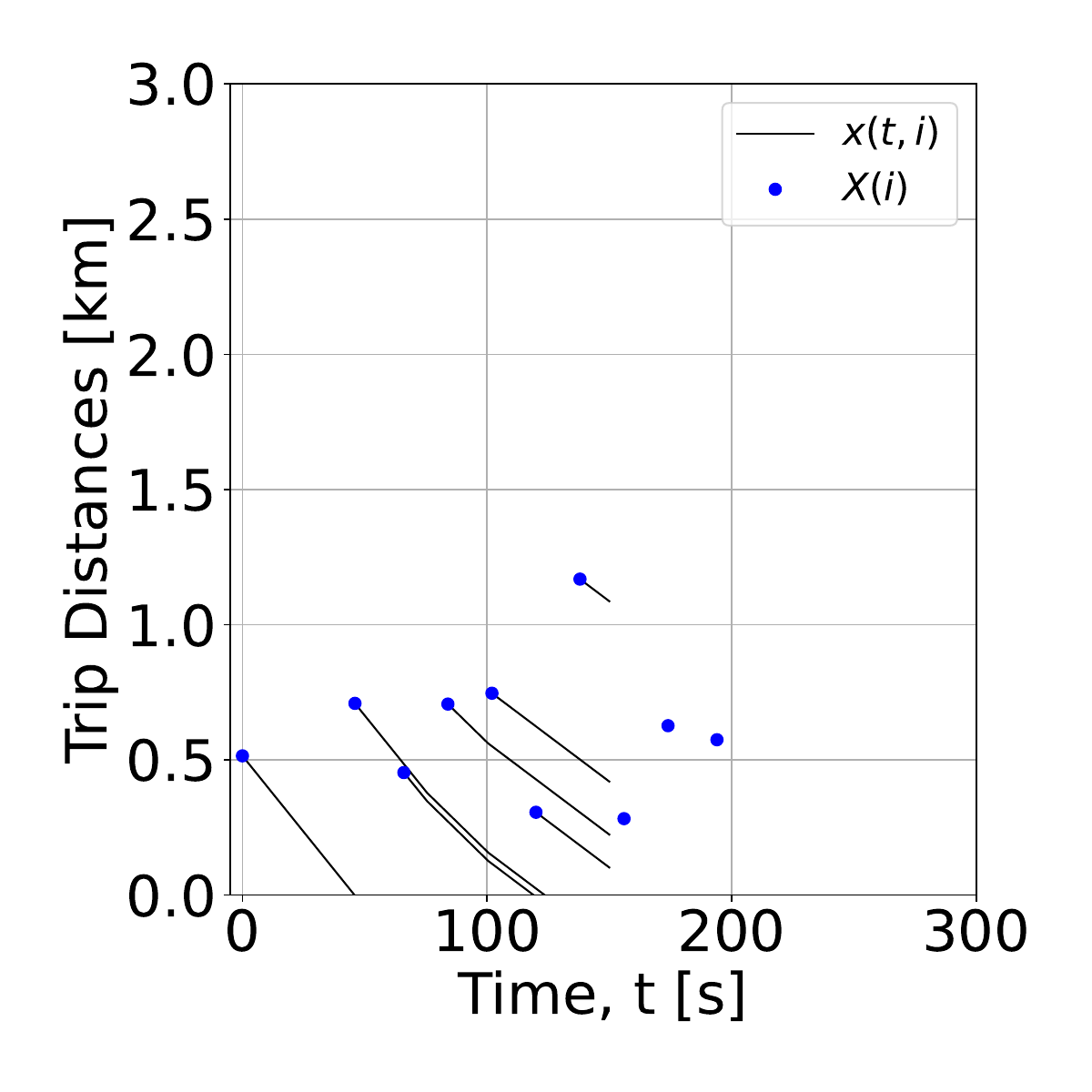}
\caption{ \small } 
\end{subfigure}%
\begin{subfigure}{.32\textwidth}
\includegraphics[width=1\textwidth]{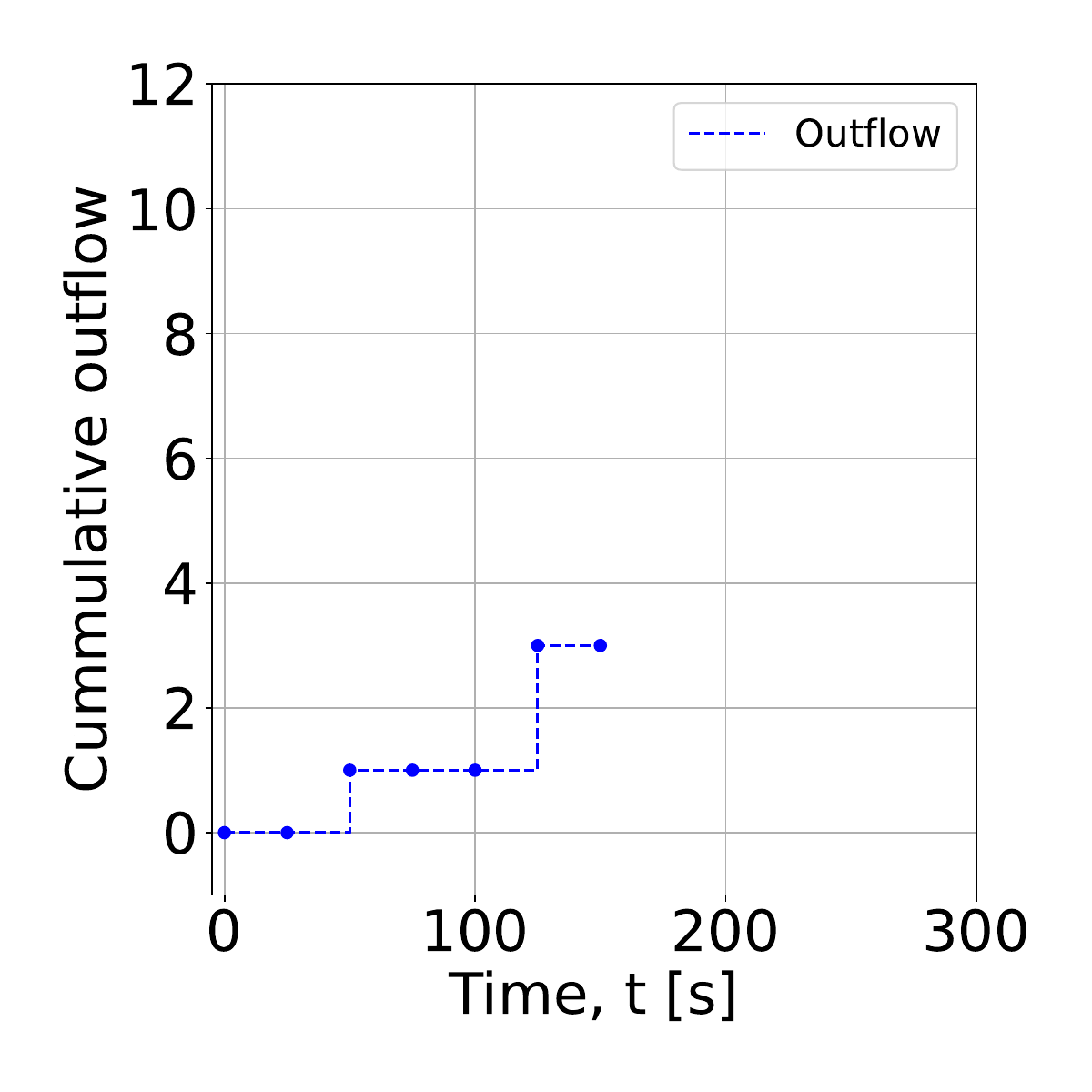}
\caption{ \small } 
\end{subfigure}%
\
\begin{subfigure}{.32\textwidth}
\includegraphics[width=1\textwidth]{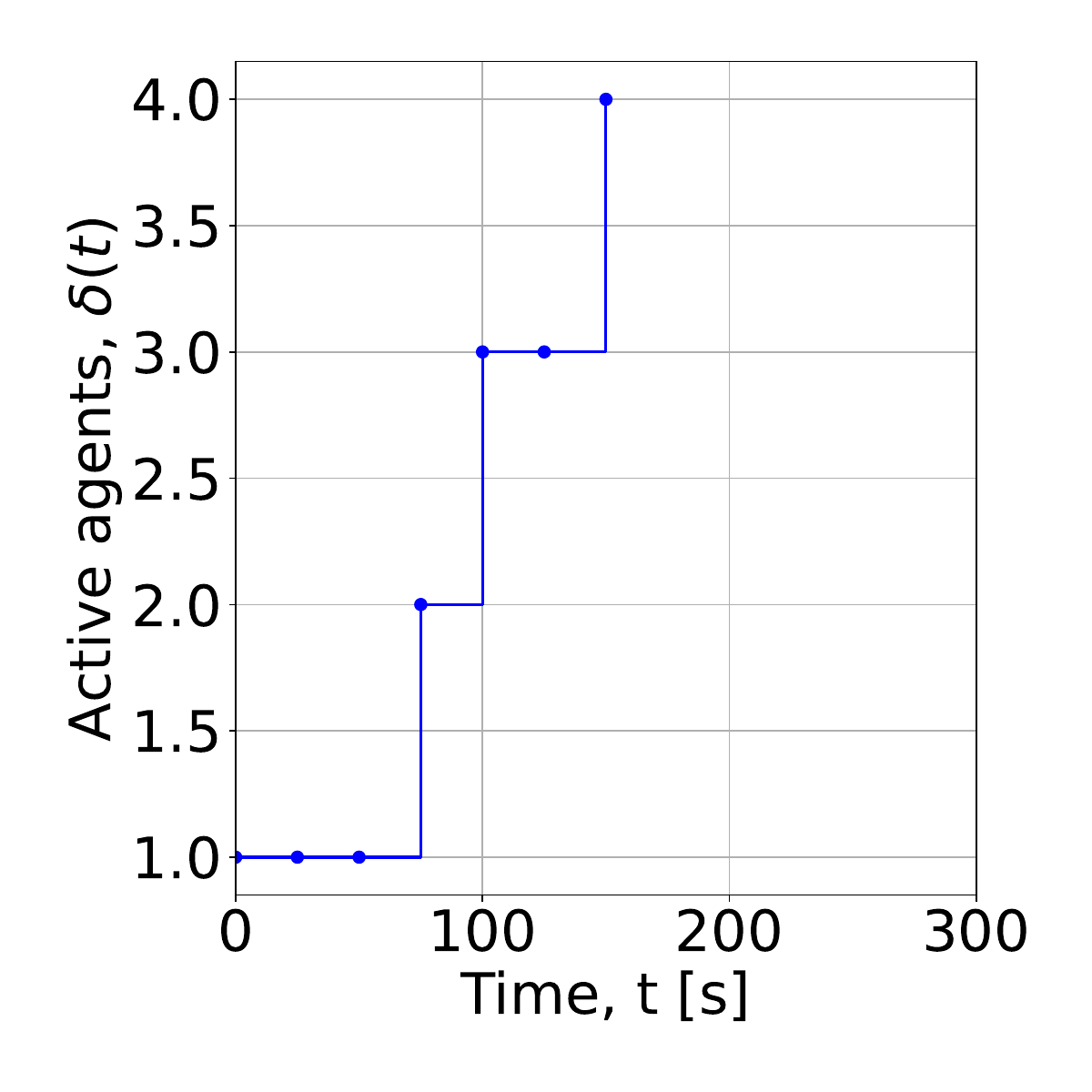}
\caption{ \small } 
\end{subfigure}%
\begin{subfigure}{.32\textwidth}
\includegraphics[width=1\textwidth]{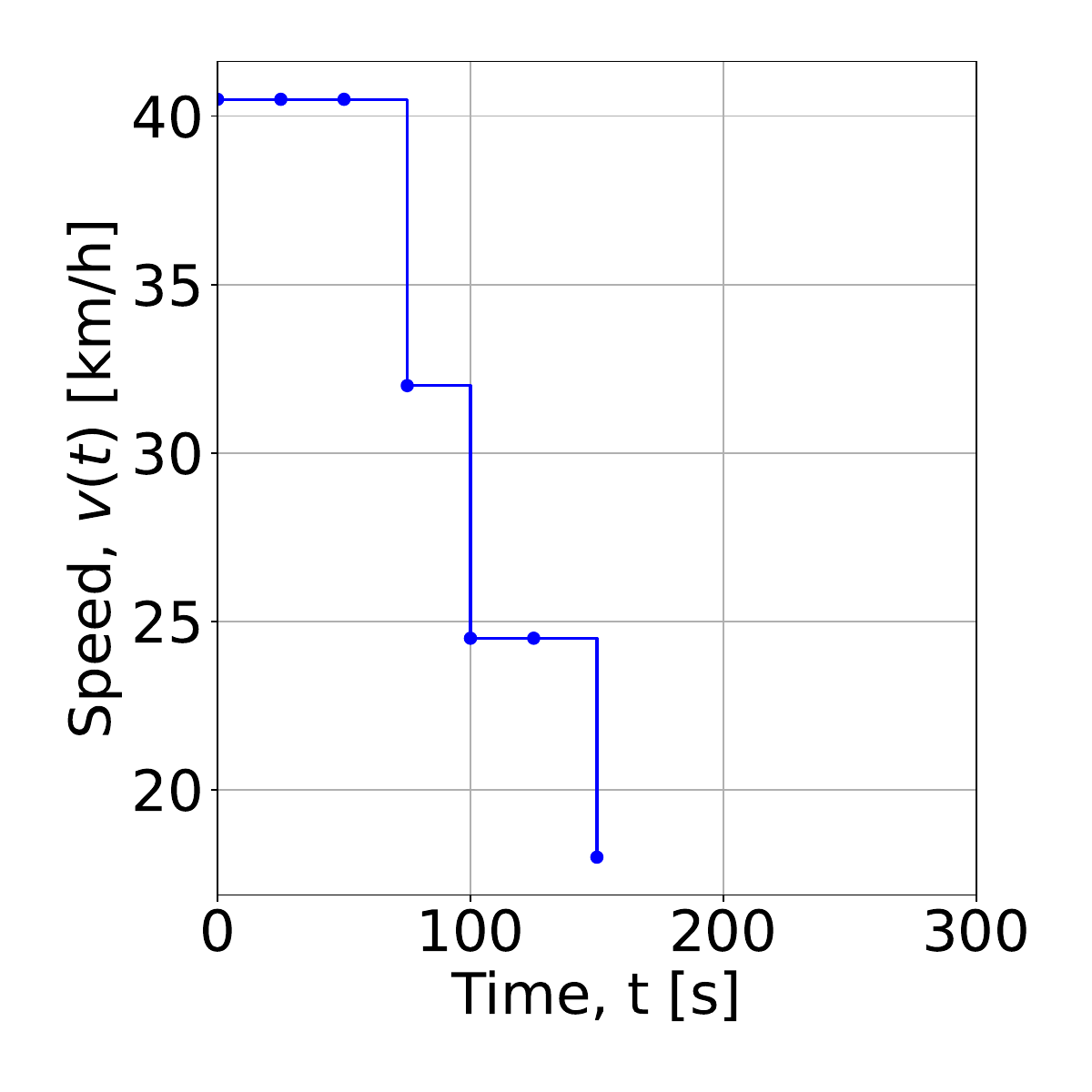}
\caption{ \small } 
\end{subfigure}%
\begin{subfigure}{.32\textwidth}
\includegraphics[width=1\textwidth]{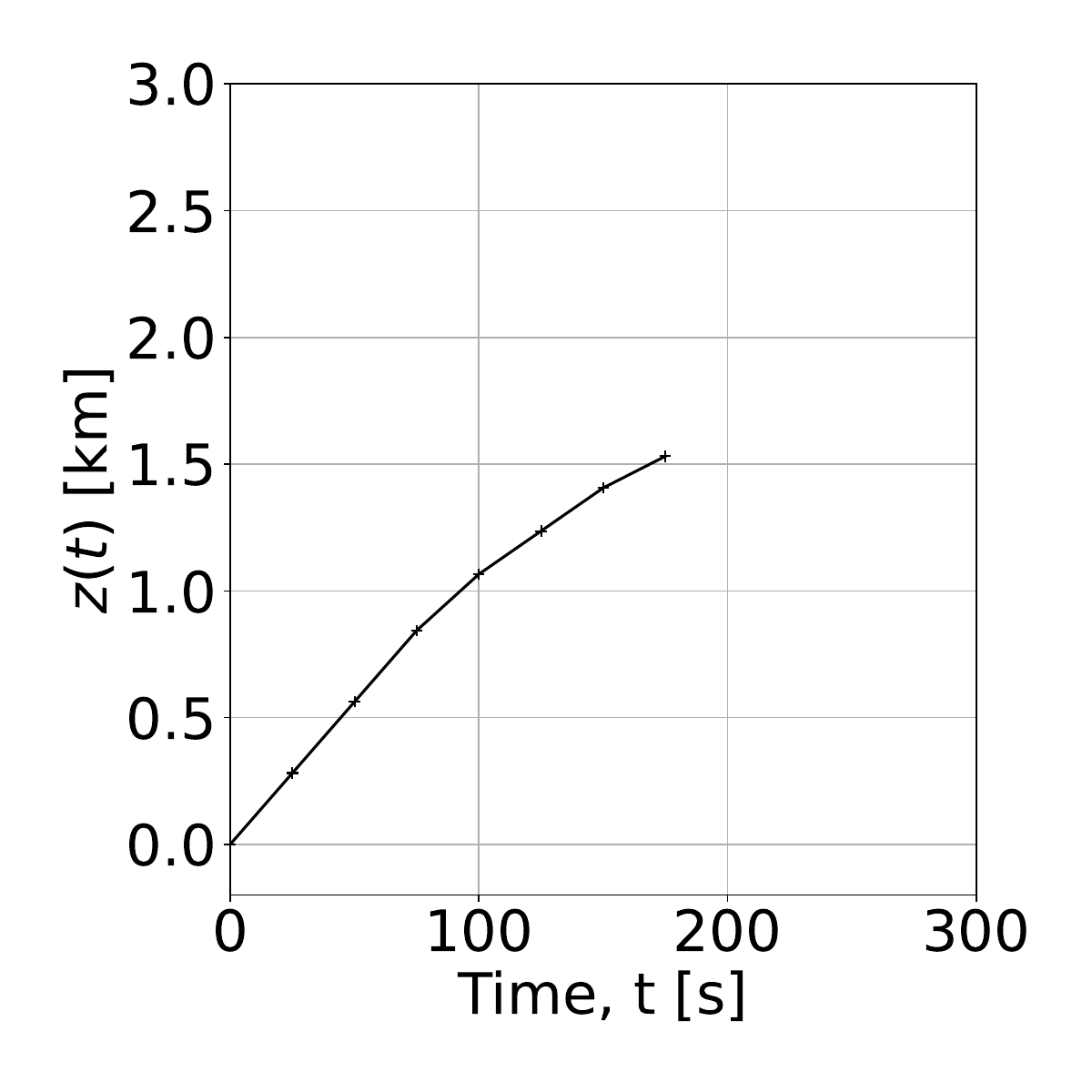}
\caption{ \small } 
\end{subfigure}%
\caption{ \small \small Example with 10 agents in 200 seconds, NFD as $V(\rho) = 50(1- \rho/10)^2$, $L_N=1$ km, $\Delta t =25$ s, simulation stopped at $t=150$ s.} \label{fig:simul_naive}
\end{figure}

The AB$^2$M consist on three main parts (see \reff{fig:overview}): Part 1 is the setup, where the agents are generated with their characteristics; Part 2 is the simulation process updating the speed of the system and the positions of the agents (i.e., the discrete trajectories of the agents in the $x-t$ plane, see \reff{fig:simul_naive}(b)); and Part 3 is the post-process, used to determine the the travel times of the agents.
As in any dynamic system, the AB$^2$M needs boundary conditions and initial conditions.
The boundary conditions are characterized by a sample of $I$ agents, that is generated by the joint distribution. 
The travel time experience of each agent may influence the demand, indicated with a dashed arrow in \reff{fig:overview} and should be studied as departure time choice \citep{Lamotte2018_Morningcommute, Ameli2022Departure}. 
However,  the influence of the travel time on the demand (e.g., departure time choice) is out of the scope of this paper, and we assume that the population size and joint distribution are exogenous and given. 

For Part 1, we consider simple  agents that only have two characteristics: the trip distance, $X(i)$, and the trip start time, $T(i)$. Thus, the model demand input can be a joint trip distance and departure time distribution, $ \tilde \varphi_{T,X}(t,x)$, \citep{jin2020generalized,Martinez2021On}. In the case that this a joint distribution is a probability distribution, the demand of the AB$^2$M can be obtained as a sample of $I$ agents from the joint probability distribution. In the AB$^2$M, the demand can be defined by agents (dots) in the $X-T$ plane. In the future, the socio-economic characteristics, or trip purpose could be added to the agents' characteristics to endogenously capture the departure time and the mode-choice, for example.  In Part 1, the agents are sorted by departure time as a fixed priority queue  and are given an index $i = {1 , ..., I}$, which can be interpreted as the trip ID, based on their start time, i.e., $T(i) \leq T(i+1)$.

In Part 2, the actual trip flow dynamics are modeled. If the study period starts at $t=0$,  w.l.o.g.,  we will assume that no agents have entered the system before that time, i.e.,   $E(t<0),z(0)=0$. 
We consider a fixed time step, i.e.,  $\Delta t^j= \Delta t$ and we divide the total simulation period $t_f$ into $K$ intervals of $\Delta t= \frac{t_f}{K}$. 
The positions of the agents and the speed of the system are updated for each time step following the six steps as shown in \reff{fig:overview}, and presented in \reff{fig:simul_naive} for a numerical example. 
The AB$^2$M can be easily extended to be an event-based simulation.

First, $E(t)$ is obtained from calculating the number of agents that have already started $T(i) < t$.
A new discrete variable  $x(t,i)$  stores the remaining trip distance to the destination for  $i$'s agent, which is updated. 
For each new trip $x(t,i)=X(i)$, and for agents that were already in the system their remaining trip distance is $x(t,i) = X(i) - \int_{T_i}^t v(\tau) d \tau$. From \refe{eq/cumtravel} we have
\begin{equation}\label{eq/x_z}
   x(t,i) = X(i) + z(T(i)) - z(t).
\end{equation}
The trip $i$ is completed when $x(t,i)=0$. Then, the total number of trips completed is 
\begin{subequations}\label{eq/g_assum}
\begin{equation}\label{eq/g_assum_a}
    G(t) = \sum_{i=1}^{E(t)} \beta_{x(t,i)},
\end{equation}
where $\beta_{x(t,i)}$ is defined as
\begin{equation}
    \beta_{x(t,i)} = \begin{cases} 1 \text{ if } x(t,i) \leq 0 \\
    0 \text{ if } x(t,i) > 0 .
    \end{cases}
\end{equation}
\end{subequations}
By conservation of vehicles, we have $\delta(t) = E(t) - G(t)$,
and the speed is obtained from the NFD, i.e., $v(t) = V \left( \rho(t) \right)$, where $\rho(t) = \frac{\delta (t)}{L_N}$ is the density. 
Finally, one updates the characteristic network traveled distance as 
\begin{equation*}
    z(t+\Delta t) = z(t) + v(t) \Delta t.
\end{equation*}

In Part 3 (the post-process), the completion time of the agents ($\hat T (i)$) is calculated by solving
\begin{equation}\label{eq:postporcess}
      X(i) = z (\hat T(i)) - z(T(i)).
\end{equation}
Then, the travel time of each user is determined as 
\begin{equation*}
    \gamma (i) = \hat{T}(i) - T(i).
\end{equation*}

The native formulation presented does not take advantage from the fact that once a trip has been completed we do not need to store its information anymore. 
A way to reduce the computational cost of the summation in \refe{eq/g_assum}, would be to use a new variable $\hat x(t,j)$, where only the active trips are sorted. Thus, the size of the variable is time dependent, i.e., $E(t)-G(t)$. However, this is still computationally expensive, because at each time step, all the remaining trip distances need to be updated.

The AB$^2$M can capture the dynamics in an exact way for a deterministic set of trips by running the simulation once. 
Further, it can be used to study a probabilistic joint distribution, $\tilde \varphi(t,x)$, including the assumption of a stochastic exogenous trip initiation rate and/or stochastic trip distance distribution. 
In that case, Monte Carlo simulations would be needed for the modeler to analyze the system.
Then, one can determine the expected behavior of the system and the standard deviation of the main variables involved. This allows to account for and study the variability of those variables, such as travel time.
It is important to note that to perform Monte Carlo simulations, it is important to ensure a low computational complexity of the AB$^2$M. 
In the following we will discuss how the AB$^2$M can become computational efficient by treating the active agents as a priority queue (Section \ref{sec:efficiency}) and through its scalability property (Section \ref{sec:scalability}).

\section{Priority Queue Formulation  of AB$^2$M}\label{sec:efficiency}

Updating and tracking the remaining trip distance of circulating trips requires high memory storage, and computational cost. 
Thus, we propose a new formulation of the AB$^2$M that is more efficient.
To do so, we define the sorted remaining trip distance as $y(t,n)$, where $n=\{1, ..., E(t)\}$ represents the trips ordered by their remaining trip distance at time $t$. We refer to $n=N(t,i)$ to the position  of trip $i$ at time $t$ in this ordered list.  Note that $y(t,n)$ increases with $n$ and is only guaranteed to decrease in $t$ for $n\leq G(t)$, since active trips could change position in the ordered list when new trips are added to the system. Then, \refe{eq/g_assum_a} can be modified to consider only the active number of trips
\begin{equation*}
    g(t) = \sum_{n=G(t)}^{E(t)} \hat \beta_{y(t,n)},
\end{equation*}
where $\hat \beta_{\hat x(t,i)}$ is defined as
\begin{equation*}
   \hat \beta_{y(t,i)} = \begin{cases} 1 \text{ if } y(t,n) \leq 0 \\
    0 \text{ if } y(t,n) > 0 .
    \end{cases}
\end{equation*}

However, FIFO does not hold for bathtub models in general. Thus, sorting the whole set of agents every time step becomes computationally very expensive.
In the following, we show how the ``shorter-(characteristic)-distance-first-out"  (SCDFO) principle \citep{jin2020generalized} enables an efficient sorting of trips.

\subsection{Shorter-Characteristic-Distance-First-Out principle }\label{sec:SCDFO}

\begin{figure}
    \centering
    \includegraphics[width=0.55\textwidth]{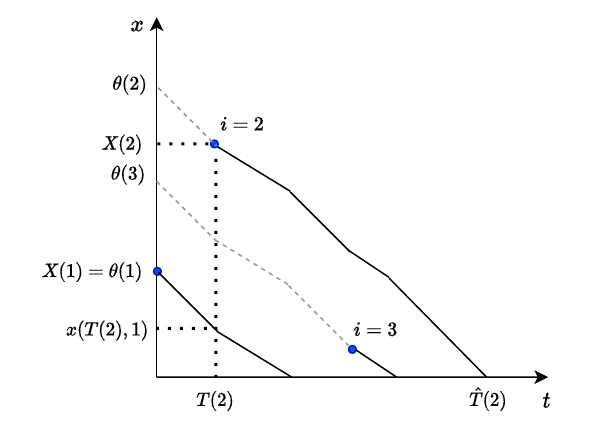}
    \caption{ \small Representation of characteristic trip distance $\theta(i)$ from the trajectories in relative space.}
    \label{fig:theta_trajectories}
\end{figure}

We extend the concept of the characteristic distance proposed by \cite{jin2020generalized} for individual agents explicitly, and we define the ``characteristic trip distance" for each agent $i$ as 
\begin{equation}\label{eq/thetai}
    \theta(i) = X(i) + z(T(i)),
\end{equation}
which is a time-independent characteristic of the agent. From  \refe{eq/x_z} and \refe{eq/thetai}, the remaining trip distance at each time instant $t>T(i)$, is related to the characteristic trip distance as
$ x(t,i) = \theta(i) - z(t)$. 
Although FIFO does not hold for bathtub models in general\footnote{For the BBM when all trips have the same distance FIFO also holds for the bathtub models.},  trips with shorter $\theta(i)$ will exit the network earlier, i.e., trip $i$ will be competed earlier than trip $j$ if $\theta(i)< \theta(j)$,
which is referred to as the SCDFO principle \citep{jin2020generalized}. 
In other words, ordering trips based on their remaining trip distance is equivalent to ordering trips by their characteristic trip distance, $\theta(i)$.
Using this property, which is the counterpart of FIFO in the absolute space, one is not required to update the remaining trip distance at each time step. Instead, the characteristic trip distance is calculated only once (when the agent enters the network) by tracing back in the relative space, see the dashed lines in \reff{fig:theta_trajectories}, where the solid lines in \reff{fig:theta_trajectories} represent the trajectories of the agents in the relative space.

As mentioned earlier, the AB$^2$M formulation is equivalent to a microscopic model (in Lagrangian coordinates) of the trip flow dynamics in the relative space. However, the vertical distance between the trajectories has no physical meaning. Because the speed of all agents is the same, the trajectories will not cross, but two agents can have overlapping trajectories if they have the same $\theta(i)=\theta(j)$ even if they start at different times $T(i) \neq T(j)$ and have different trip distances $X(i) \neq X(j)$.

The advantage of SCDFO is that the agents can be sorted by increasing $\theta(i)$ to facilitate the calculation of $G(t)$. Moreover, from \refe{eq:postporcess} and \refe{eq/thetai} the post-process can be simplified to solve $ \theta(i) = z(\hat{T}(i))$.

\subsection{Priority queue definition and efficient algorithm}\label{sec:algorithm}

To take advantage of the SCDFO principle, we propose a new main variable, which is an ordered list of characteristic trip distances of the circulating agents $\Theta(t,n)$. This vector satisfies $\Theta(t,N(t,i))= \theta(i)$, where $N(t,i)$ is the position in this ordered list of the trip $i$ at time $t$. Thus, $\Theta(t,N(t,i)=1)$ is the active trip at time $t$ with the shortest characteristic trip distance, and $\Theta(t,\delta(t))$ is the active trip with the longest characteristic trip distance.

In this efficient formulation, Steps 2 and 3 of Part 2 (\reff{fig:overview}) are modified.
Instead of updating the agents' remaining trip distance, we update $\Theta(t,n)$ in Step 2. 
Since the characteristic trip distance is a time-invariant feature of each agent, the modeler only needs to sort the new arriving trips into an already existing ordered list, which represents a significant computational complexity reduction. 
Then, the completion of trips, $g(t)$, corresponds to the sum of 
all agents with $\Theta(t,n) \leq z(t)$, i.e., 
\begin{equation}\label{eq/ABBM}
    g(t) = \max_n \left\lbrace \Theta (t,n) \leq z(t) \right\rbrace.
\end{equation}

The position of trip $i$ in the ordered list by characteristic trip distance may change with time, since trips with start time $T(j)>t$ might have shorter characteristic trip distance than existing trips, i.e.,  $\theta(j) < \Theta(t, \delta(t))$. Thus, $N(t,i)$ is time-dependent for a given active trip $i$.

\begin{figure}
  \centering
  \includegraphics[width=0.75\linewidth]{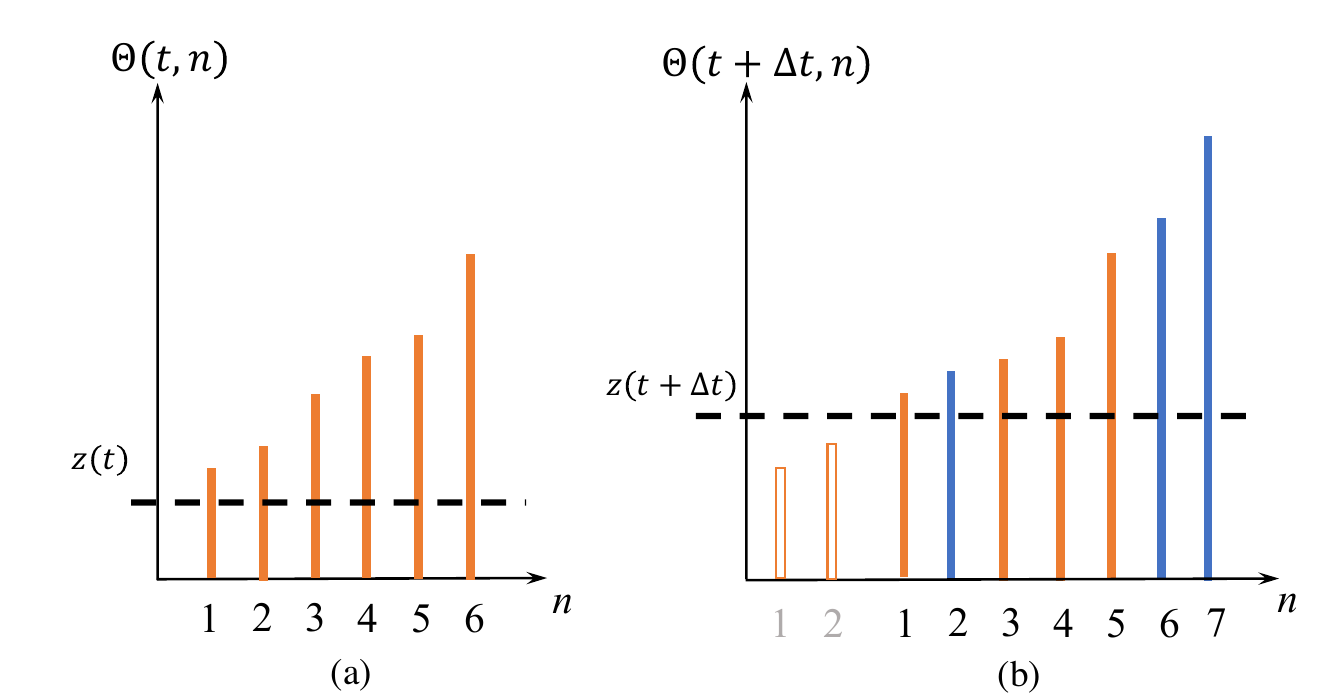}
    \caption{ \small Evolution of $\Theta(t,x)$. When the characteristic trip distance of the new trips is the largest of all the current circulating agents, it is added at the end of the queue. If a new trip $\theta(i)$ is not the longest, it is added between active trips.}
    \label{fig/theta/cont}
\end{figure}

By definition of ordered list, $\Theta (t,n) $ is monotonically non-decreasing with $n$.
In contrast, its time dependency is non-monotonic for a given $n$. 
The discrete interpretation is presented in \reff{fig/theta/cont}, which shows the agents that initiate their trip between $t$ and $ t+ \Delta t$ in blue, and the agents that have completed their trip  outlined.

In the following, we discuss the details of the fixed time-step  leveraging the priority queue formulation, i.e., Algorithm \ref{Algorithm_heap}.
Since $\Theta(t,n)$ is a priority queue, we propose to use an ordered list using min-heaps to store $\Theta(t,n)$ in the algorithm. 
Heaps are a type of data structure, and a min-heap is a binary tree where the root parent node is smaller than the children. Heaps are usually implemented to handle priority queues and are common in shortest path algorithms \citep{JOHNSON1975Priority}. Min-heaps also allow an easy restructure when new elements are added or when certain elements are removed from the list. 
Min-heaps are defined by two characteristics: (i) only elements on the top (i.e., shorter characteristic trip distance) are removed, and (ii) new trips can be inserted in any place, since they can have a shorter or longer effective distance than active trips in the network.

\begin{algorithm}
\small
\SetAlgoLined
 \% Initialization of variables\\
 $t=0$, $E(t)=0$, $z(t)=0$\;
 \% Simulation \\
 \For{ $t= 0$ to $t=t_f$}{
  Obtain $E(t)$ from $T(i)$; 
 \If{$E(t) > E(t-\Delta t)$}{
    \For{$j$ = 1 to $j$= $E(t) - E(t-\Delta t)$}{
    $\theta(E(t)+j) = z(t) + X({E(t)+j}) - (t-T(i))v(t-\Delta t)$;\\
    Insert $\theta(E(t) + j)$ in an ordered way in heap $\Theta(n)$; \% Cost $\mathcal{O}(\log n)$\\
        }
    }
    $g(t)=0$;\\
 \While{$ \Theta(n=1) \leq z(t)$}{
  Drop $ \Theta( n=1)$ from the heap; \% Cost $\mathcal{O}(\log n)$\\
  $g(t)= g(t) +1$;
  }
  $G(t)=G(t-\Delta t) +g(t)$;  \% Calculate cumulative outflow \\
  $v(t) = V((E(t) - G(t))/L_N)$; \% Update of speed \\
  $z(t+\Delta t) = z(t) + \Delta t v(t)$;\\
  $t= t + \Delta t$
 }
 \caption{ \small \small Pseudo-code of the proposed algorithm with fixed time step.}\label{Algorithm_heap}
\end{algorithm}

In Algorithm \ref{Algorithm_heap}, in each time step the cumulative inflow is calculated.
If $E(t) > E(t-\Delta t)$, new trips are entering the system, and their characteristic trip distance should be calculated from \refe{eq/thetai}. 
Then,  the entering trips' $\theta(i)$ are  introduced  into $\Theta(t,n)$. 
The completion rate of trips  \refe{eq/ABBM}, can be determined by comparing the first element of $\Theta(t,n)$ with the characteristic network traveled distance. If $\Theta(t,1) \leq z(t)$, the trip is removed and $\Theta(t,n)$ is updated. 
This process is iterated until all active trips have a characteristic trip distance larger than $z(t)$, i.e., 
$\Theta(t,1)>z(t)$. This is only efficient if there are very few active trips. Otherwise, it would be worth to use a bisection method to find the $\Theta(t,n)$ closest to $z(t)$.

\section{Scalability Property of the Bathtub Models}\label{sec:scalability}

A promising direction to make AB$^2$M more efficient is to explore the concept of ``downscaling'' already used to reduce the computational cost in traditional ABM in the absolute space \citep{Nicolai2012Using}. There are two approaches to downscaling: (i) use so-called ``super-agents" (SA) that represents several agents, although it requires a re-programming of the agents and their behaviors, and (ii) the so-called ``one-represents-several" (ORS) methods, where the network characteristics (such as capacity and jam density) are reduced based on scaling ratios (usually proportional to the reduction of the number of agents). The downscaling of agents in MATSim has been widely implemented, using the ORS method by downscaling linearly the capacity of the links as $C_r=C r$ and nonlinearly the jam density of the links as $\rho_{j,r} = \rho_j r^{0.75}$, see details in \citep{Nicolai2012Using}.  The scaling factor, $r$, ranges from 1\% in Munich; see \cite{KICKHOFER2015Pricing} to 25\% in Dublin; see \cite{Mcardle2014Using}. 
Despite the long-term implementation of this approach, there are recent studies that conclude that such downscaling might introduce bias into the results \citep{LLORCA2019Effects, BENDOR2021Population}.
\cite{LLORCA2019Effects} performed simulations on the Munich network and showed how the average travel time is highly dependent on the scaling factor. Further, \cite{BENDOR2021Population} analyzed the effect of downscaling on the accuracy of the results at a simple network (Sioux Falls). Both studies concluded that downscaling below $r=0.1$ might introduce bias into the results.
Moreover, there is a physical limit for downscaling ABM in absolute space. Since the characteristics of the individual links are scaled, the links cannot be reduced to fit less than one vehicle. The theoretical lowest jam density that can be considered is $\rho_{j,r}=1$ veh/km/lane, which limits the scaling factor $r$.

The downscaling for bathtub models is a natural idea that has been used by several researchers but not systematically discussed in the literature. For example, \cite{Lamotte2018} propose to use weights for trips with different characteristics and then calculate the speed as $v=V\left( \frac{\sum_i \omega_i \delta_i}{L_N}\right)$. Note that their approach does not involve downscaling the system (supply), and the network lane miles are not modified. 

This section fills this gap by methodically studying the scalability (including both downscaling and up-scaling) in the relative space. First, identify two types of scaling strategies for bathtub models in  \ref{sec:downscaling_prop} and  later discuss the numerical errors that can be introduced by downscaling in Section \ref{sec:downscaling_error}.

\subsection{Scaling of bathtub models}\label{sec:downscaling_prop}

When considering the GBM \refe{eq:gbm}, we can observe that the dynamic system can be scaled by $r$ as
\begin{equation}
        \frac{\partial K(t,x) \cdot r }{\partial t} - V\left(\frac{K(t,0) \cdot r}{L_N \cdot r} \right) \frac{\partial K(t,x) \cdot r}{\partial x} = e(t) \tilde \Phi (t,x) \cdot r ,
\end{equation}
where $r$ is the scaling factor. If $r<1$, the system is down-scaled, and if $r>1$, the system is up-scaled. In the scaled system the time and speed are invariant. Since the speed depends on the density through the NFD, the density in the scaled system should be invariant too. Therefore, a good way to look at the GBM to assess its scalability is to look at the density evolution, which in turn defines the speed evolution.  From \refe{eq/e} the GBM \citep{jin2020generalized}  can be written as 
\begin{equation}\label{eq/scalab}
    \dot \rho(t) = \frac{I}{L_N} \int_0^{\infty} \tilde \varphi(t,x) dx   - \rho(t)  \varphi(t,0) V (\rho(t) ).
\end{equation}

A first and natural way to downscaling is to consider that the scaling factor $r$ is used to modify the trip initiation rate $e(t)$, the length of the network, $L_N$, and the number of active trips with remaining distance not smaller than $x$, i.e., $K(t,x)$. 
We refer to this strategy as flow-based scaling. 
Clearly, the traffic dynamics described by \refe{eq/scalab} on multiple neighborhoods or cities will be the same as long  these have the same NFD, $\tilde \varphi_{T,X}(t,x)$, and ratio $\frac{I}{L_N}$. 
Therefore, the flow-based scaling modifies the supply and demand without introducing biases. On one hand, the demand  is scaled  through the number of agents $r I$, without changing $\tilde \varphi(t,x)$.  On the other hand, the supply is scaled by changing the total network lane distance, i.e.,  $r L_N$, without changing the parameters of the NFD.

Since the demand is defined by total vehicle-distance units, it may be interesting to explore another (not so natural) type of scaling, by modifying the distance traveled instead of the number of vehicles. In this case, the supply is scaled in the same way, i.e.,  modifying the network length, $L_{N,r} = r L_N$. However, the demand is scaled maintaining the number of agents, and modifying the average distance of those agents, as $\tilde D_r(t) = r \tilde D(t)$.
This will be referred as distance-based scaling, and may introduce some bias during the transition states.

Although the impact of distance-based scaling can not be neatly represented in \refe{eq/scalab}, it can be shown that during steady states\footnote{ In this case we refer to the definition by \cite{LAVAL2023Effect}: ``In steady-state [...] the circulating flow over the total network distance, has to match the incoming production [of trip distance]'', i.e., $\dot m (t) =0$.}, the systems are equivalent, with both flow-based and distance-based scalings. From \refe{eq-def-m} and $\dot m (t) =0$ we have
\begin{equation}
    \frac{e(t) \tilde D(t)}{L_N} = \rho(t) V \left( \frac{\delta(t)}{L_N} \right).
\end{equation}
Whether the same steady state is stable, reachable, etc. for different demand patterns is a relevant research question out of the scope of this paper. 

As discussed earlier, there is a physical limit to downscaling the model in the absolute space.
In contrast, in relative space there is no such limit, and one can simulate 1.25 million agents on a network of $L_N=1000$ km (computational cost 2.2s) with 1250 agents on a network of ${L}_{N,r}=1$ km (0.12s), which represents a downscaling of 0.1\% in the number of agents and a computational cost reduction to $\sim$5\%. A detailed discussion on the computational complexity of the algorithm and benefits of downscaling will be presented in Section \ref{sec:complexity}. 
However, the scaling in the relative space demand should be done carefully, since there is a risk that downscaling the network too much introduces numerical errors, as will be discussed in Section \ref{sec:downscaling_error}. 

This unbiased flow-scaling property of the bathtub model allows the modeler to study the traffic dynamics of a large system with a lower number of agents and a smaller network with $r<1$. 
For example, the modeler can create a normalized twin city with $L_{N,r}=1$ and total demand $I_r = \frac{I}{L_N}$ to study traffic patterns. This scalable property of the bathtub model will allow modelers to use non-dimensional analysis to study the traffic congestion patterns in very large cities. 
Alternatively, if the modeler has information on the characteristics of a reduced number of agents in the system, $I_r$, instead of the total number of trips during a study period in a study network and mode ($I$) on a network of $L_N$ lane-distance, the traffic dynamics in the system can be modeled with agents $I_r$ in an equivalent city setting the network lane distance to $\frac{I_r}{I} \cdot L_N$.


\subsection{Numerical errors induced by flow-based scaling in the AB$^2$M}\label{sec:downscaling_error}

\begin{figure}
  \centering
  \includegraphics[width=\linewidth]{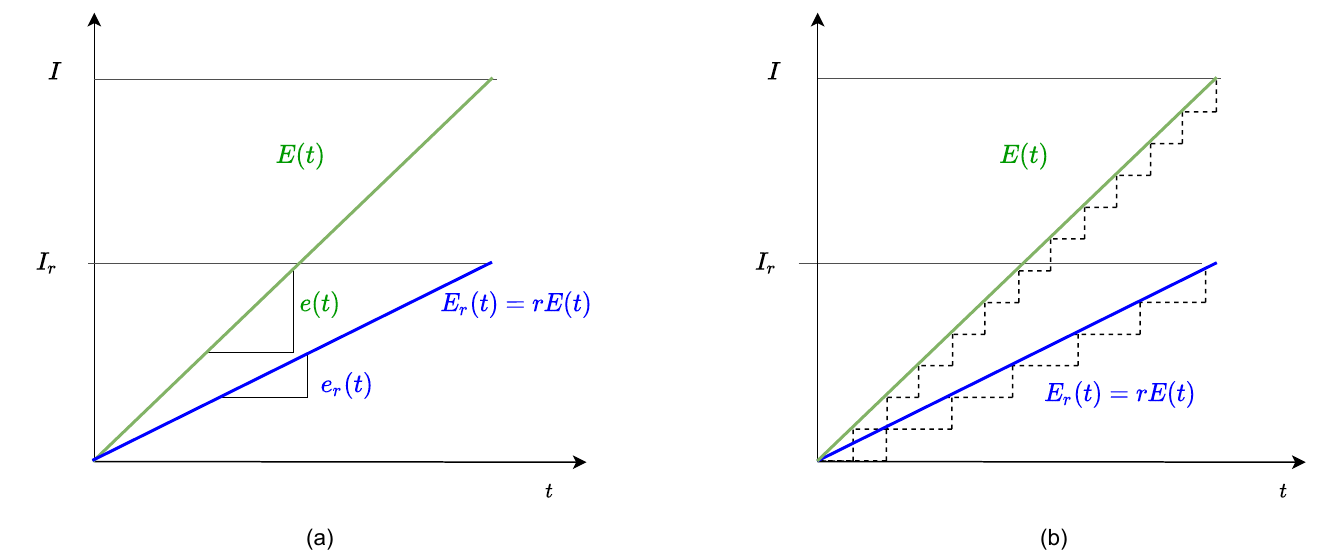}
    \caption{downscaling of demand: (a) For continuous $E(t)$. (b) For piece-wise constant $E(t)$.}
    \label{feig:normalization}
\end{figure}

From \refe{eq/e}, the scaling of the total number of agents influences the inflow of trips as $e_r(t) = I_r \tilde \varphi(t)$.
For continuous demand, the cumulative trip initiation rate is $E_r(t)=r E(t)$. However, since the AB$^2$M is by nature discrete, the modeler should first approximate the piecewise constant $E(t)$ to a piecewise linear function, scale the approximation down, and discretize it again as a piecewise constant function; see Figure \ref{feig:normalization}.

There are two different types of error that can arise when downscaling the AB$^2$M: First, the downscaling can lead to non-integer number of agents in a discrete demand, as discussed in Section \ref{sec:downscaling_discrete}. Second, the downscaling can reduce the network in a way where a single agent starting or finishing their trip leads to large speed variations (and non-representative results of the original system), as will be discussed in Section \ref{sec:speed_variation}.

\subsubsection{Discrete and deterministic demands}\label{sec:downscaling_discrete}

\begin{figure}
    \centering
    \includegraphics[width=\textwidth]{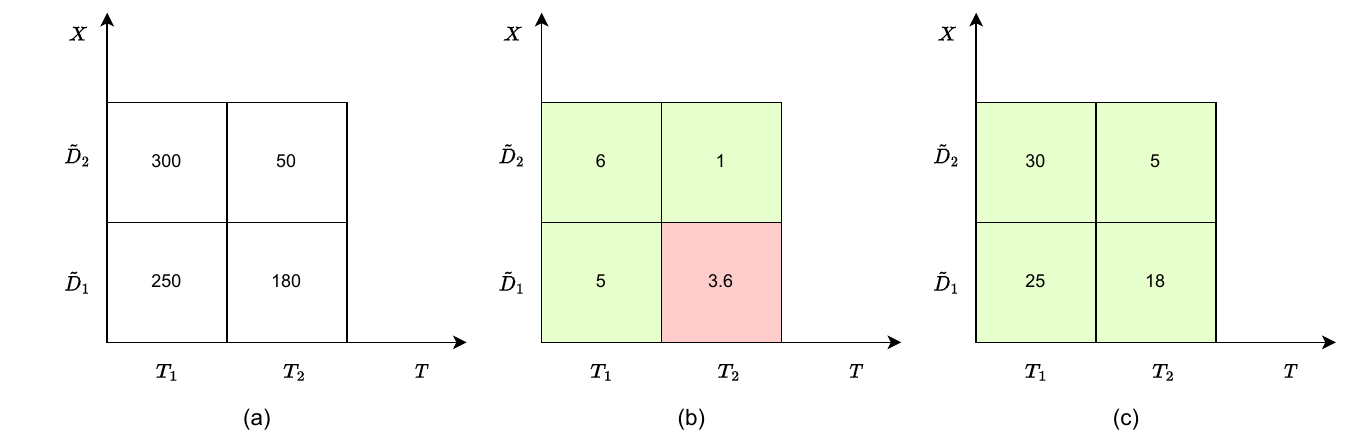}
    \caption{\small Example of ensuring integer number of agents with deterministic discrete demands. (a) Original sample of agents, (b) Scaled demand with $r=1/50$, (c) Scaled demand with $r=1/10$.}
    \label{fig:scaling_determ}
\end{figure}

In the case of a discrete demand, where $I \tilde \varphi(t,x)$ is defined by an integer number of agents that start their trip at certain discrete times, $T(i)$, with certain trip distances $X(i)$, the downscaling can introduce numerical errors if $r$ is not chosen carefully. The reason is that not all agents might be represented if downscaling requires rounding. Let us consider a discrete demand, where $n_{km}$ is the number of agents with departure times $T_k$ and trip distances $X_m$. 
In that case, the lowest scaling ratio $r_{min}$, needs to guarantee that $r_{min} \cdot n_{km}$ are integers $\forall m, k$. Therefore, the lowest scaling ratio can be obtained from $\frac{1}{r_{min}}=GCD$, where $GCD$ is the highest common divisor of all $n_{km}$. 
For example, a network with two trip distances $\tilde D_1$ and $\tilde D_2$ and two departure times $T_1$ and $T_2>T_1$. As depicted in Figure \ref{fig:scaling_determ}(a), the demand is defined as follows: at time $T_1$, 550 agents start their trip, with 250 agents having $\tilde D_1$, and 300 having $\tilde D_2$. Then, 180 agents with $\tilde D_1$ and 50 agents with $\tilde D_2$ depart at $T_2$. 
Choosing $r=1/50$ leads to a non-integer number of agents, see Figure \ref{fig:scaling_determ}(b). Therefore, choosing 3 or 4 agents with trip distance $\tilde D_1$ departing at $T_2$ would lead to numerical errors. 
Instead, the lowest scaling ratio should be $r_{min} = 1/10$, because 10 is $GCD$ of $n_{km}$, see Figure \ref{fig:scaling_determ}(c). 

In contrast, there is no such limit when the demand is stochastic. Instead, the scaling ratio $r$ will influence the sample size of agents obtained from the stochastic joint distribution of the demand. 
Then, the modeler can analyze the expected behavior through Monte Carlo  simulations. 
The scaling ratio $r$ does not influence expected behavior as long as the reduced number of agents does not induce numerical errors, as we will discuss in the next section.

\subsubsection{Speed variation}\label{sec:speed_variation}

If $r$ is too small, a single vehicle entering or leaving the system could cause significant changes in density and speed, introducing numerical errors. 
For example, modeling a network as a non-dimensional system with $r=\frac{1}{L_N}$ might not allow capturing a smooth change in the speed over time even if the time step is very small and ensures a low number of entering agents each time step. This is because a single vehicle entering or leaving the network can substantially modify speed. In fact, the rate of change in speed depends on the change in the accumulation of active trips $\delta(t)$ as
\begin{equation}\label{dvddelta}
    \frac{\text{d} V(\delta)}{\text{d} \delta} = \frac{\text{d} V(\rho)}{\text{d} \rho} \frac{1}{r L_N},
\end{equation}
which is inversely proportional to the total network length for a given density.

\begin{figure}
    \centering
\begin{subfigure}{.37\textwidth}
  \centering
  \includegraphics[width=1\linewidth]{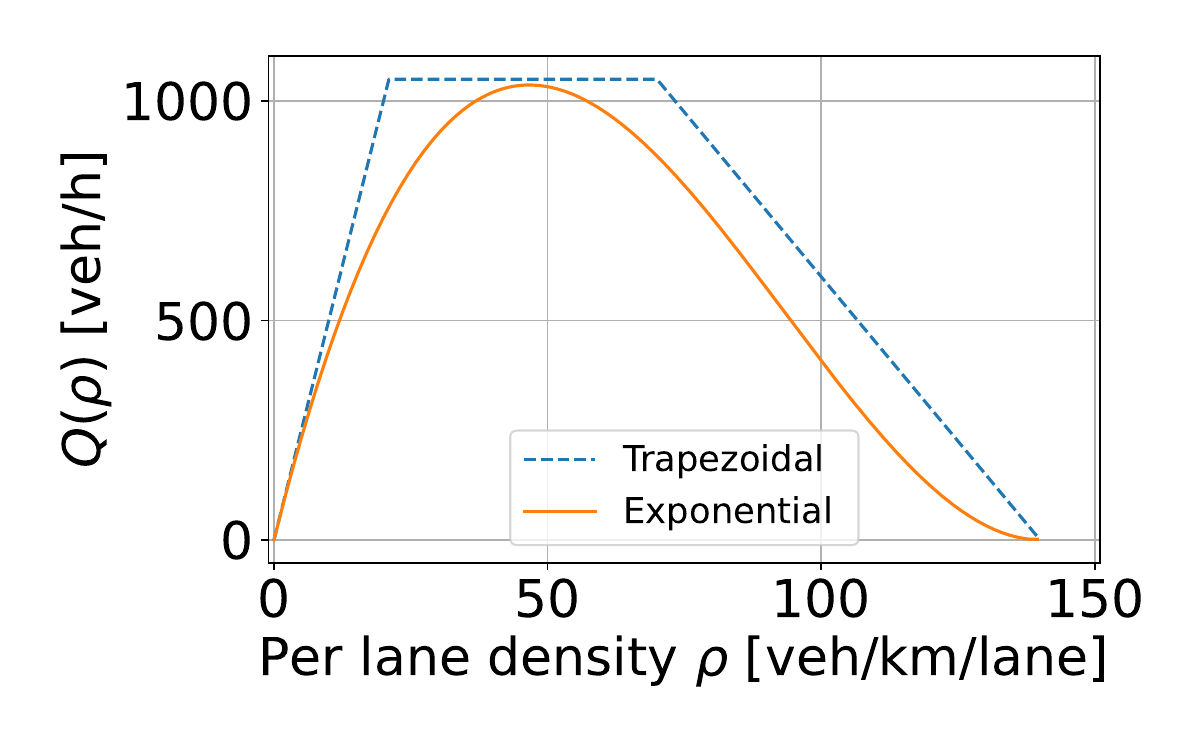}
  \caption{} \label{example:NFD}
\end{subfigure}%
\begin{subfigure}{.37\textwidth}
  \centering
  \includegraphics[width=1\linewidth]{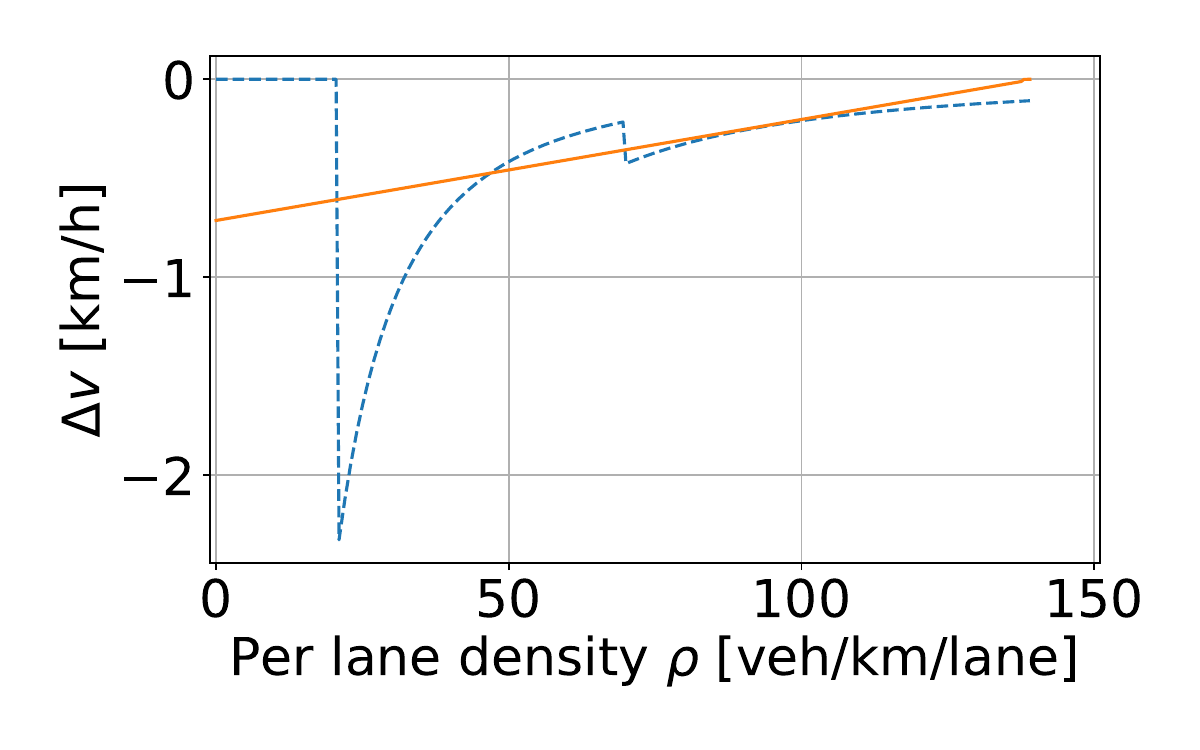}
  \caption{} 
\end{subfigure}%
    \caption{\small Numerical examples with free-flow speed of $u_f=$50 km/h and per lane jam density of $\rho_j=$140 veh/km/lane. (a) Speed density relation for the trapezoidal NFD with $C$=1050 veh/h and shock wave speed $w=15$ km/h, and the exponential NFD. (b) Speed variation for the two NFDs at different densities corresponds to a network with $L_N=1$ and a single vehicle entering or leaving the system, i.e., $\Delta \delta = \Delta \rho =1$ }
    \label{fig:dvdrho}
\end{figure}

As an example, let us consider two different NFDs presented in Figure \ref{fig:dvdrho}(a) with the same free-flow speed and per lane jam density: the trapezoidal NFD $V(\rho)= \min\{ u_f ; \frac{C}{\rho} \}; w(\frac{\rho_j}{\rho} - 1)$; and the exponential NFD $V(\rho) =u_f (1 - \frac{\rho}{\rho_j})^2$. The derivative of the speed with respect to the density is depicted in Figure \ref{fig:dvdrho}(b) for the NFDs considered. For trapezoidal NFD, the largest $\lvert \frac{\text{d}V}{\text{d}\rho} \rvert$ is observed at the critical density; while in exponential NFD the largest variation is for $\rho=0$. From \refe{dvddelta} a downscaled city with $L_{N,r} <1$ will lead to higher speed variations than those presented in Figure \ref{fig:dvdrho}(b). 
Therefore, if the modeler wants to guarantee a smooth speed (low $\Delta v$), she can determine the minimum reduced network lane distance from \reff{fig:dvdrho} as $r_{min} L_N= \frac{1}{\Delta v} \max_\rho \{ \frac{\text{d} V(\rho)}{\text{d} \rho} \}$. For example, for the above NFDs with $\Delta v=0.1$ we would have $r_{min} L_N = 10$ km for the exponential NFD, and $r_{min} L_N = 25$ km for the Trapezoidal NFD.

In summary, although there are significant computational benefits in reducing $r$, this also leads to an increase in maximum speed variation for each unit vehicle that enters or exits the system \refe{dvddelta}. Therefore, the modeler needs to evaluate this type of trade-off to choose $r$ to obtain accurate results in a computationally efficient manner. 

\section{Complexity analysis}\label{sec:complexity}


In this section, we compare the computational complexity of the algorithms proposed and analyze the impact of downscaling. We are interested in analyzing the complexity of a system with a certain simulation period of $t_f$ with fixed time step $\Delta t$. To do so, we define the average inflow rate per unit distance during the simulation period as
\begin{equation}
    \bar e = \frac{I}{L_N t_f}.\label{eq:e_average}
\end{equation}

As discussed earlier, Part 1 and Part 3 of \reff{fig:overview} are the same for the  naive AB$^2$M formulation and the efficient priority queue formulation proposed in Section \ref{sec:efficiency}. In the setup, sorting all the agents by departure time can be done with efficient algorithms relying on tree-sorting, where the cost is $n log(n)$ for a list of $n$ elements \citep{Hetland2010}. The post-process, i.e., finding the completion time for each trip, can be done numerically with any zero of functions algorithm search \refe{eq:postporcess}. For example, a binary search algorithm can be used, where the cost is logarithmic with the number of time steps \citep{Hetland2010}. 

\begin{lemma}
    The upper bound complexity of Part 1 and Part 3 of the algorithms to model AB$^2$M are $\mathcal{O} \left( I log(I) \right)$ for Part 1,  and  $\mathcal{O} \left( I log (\frac{I}{\bar e \Delta t L_N} ) \right)$ for Part 3, where $\bar e$ is the average inflow.\label{theorem1}
\end{lemma}

\begin{proof}
    The generation of $I$ agents has cost $I$ and sorting the agents by departure time can be done in $\mathcal{O} \left( I log(I) \right)$, with the most efficient algorithms relying on tree-sorting  \citep{Hetland2010}. Thus, overall, the complexity of Part 1 is $\mathcal{O} \left( I log(I) \right)$. For Part 3, the time of completion for each trip has computational upper bound complexity of $\mathcal{O}(\log \frac{t_f}{\Delta t})$, if the time step is fixed $\Delta t$. Then, the total post-process cost is $\mathcal{O}(I \log \frac{t_f}{\Delta t})$. 
    From \refe{eq:e_average}, we have the upper bound complexity for Part 3 is $\mathcal{O} \left( I log (\frac{I}{\bar e \Delta t L_N} ) \right)$.
\end{proof}

From Theorem \ref{theorem1}, the computational cost increases loglinearly with increasing the number of agents. Instead, it decreases logarithmically with larger $\Delta t$, $\bar e$, and networks. Although the last two seem counter-intuitive, with a fixed number of agents $I$ lowering the average trip initiation rate will lead to a longer simulation period and thus will increase the computational cost.

In the following, we discuss the computational complexity for the simulation dynamics (i.e., Part 2 in \reff{fig:overview}).

\begin{theorem} The naive algorithm proposed in \reff{fig:overview} has the following upper bound complexity for Part 2 of $\mathcal{O} \left(\frac{I^2}{\bar e \Delta t L_N} \right)$. This quadratic complexity from Part 2 dominates the overall complexity of the naive algorithm. \label{theorem2}
\end{theorem}

\begin{proof}
    The simulation runs for $\frac{t_f}{\Delta t}$ time steps. In each time step, the cumulative inflow $E(t)$ is obtained from a binary search among all the sorted agents, i.e., $\mathcal{O} \left( log (I) \right)$. In each time step, there is an update of the remaining trip distance as well as a search over all agents that have left their origin, $E(t)$, see \refe{eq/g_assum}. Thus, the total cost is $\mathcal{O} \left(\frac{t_f}{\Delta t} ( log(I) + I ) \right)$, and from \refe{eq:e_average}, the upper bound is $\mathcal{O} \left(\frac{I^2}{\bar e \Delta t L_N} \right)$. 
\end{proof}

\begin{theorem} The priority queue Algorithm \ref{Algorithm_heap} upper bound complexity for Part 2 is 
\begin{equation*}
\mathcal{O} \left(  \frac{I}{\Delta t \bar e L_N} \log (I) +  I  \cdot  \log (\rho_j L_N )   \right) .
\end{equation*}
Thus, the Algorithm is dominated by Part 1 computational complexity unless $\Delta t << \frac{1}{\bar e L_N}$, which would make $\mathcal{O} \left(  \frac{I \log (I)}{\Delta t \bar e L_N}  \right) $ dominate the algorithm. \label{theorem3}
\end{theorem}

\begin{proof}
    Similar to in the naive formulation, there is a cost of defining $E(t)$ for all time steps, i.e.,  from \refe{eq/g_assum} $\mathcal{O} \left(\frac{I}{\Delta t \bar e L_N} ( log(I) ) \right)$. The complexity from inserting an agent to and removing it from  the priority queue of size $\delta(t)$ is $\mathcal{O} \left( log(\delta(t) ) \right)$ for a \emph{heap} structure \citep{Hetland2010}. Thus, the computational cost of incorporating $e(t)$ elements into a sorted \textit{heap} is 
\begin{equation*}
   \sum_{k=0}^{e(t)-1} \mathcal{O}\left( \log (\delta(t-\Delta t) + k)\right).
\end{equation*}
    If $e(t) < \delta(t-\Delta t)$, it can be approximated by $\mathcal{O}\left(e(t) \log(\delta (t-\Delta t) + \frac{e(t)}{2})\right)$, which is dominated by $\mathcal{O}\left(e(t) \log(\delta (t-\Delta t))\right)$. Since the number of active trips is always  $\delta(t)  \leq \rho_j L_N$, the upper bound complexity of inserting and removing all agents is $\mathcal{O} \left( I  \cdot  \log (\rho_j L_N )  \right)$. 
\end{proof}

\begin{corollary}
Adding and removing agents to the priority queue is dominated by defining $E(t)$, when  $I >> \rho_j L_N $, e.g., for longer periods of time or high $\bar e$. Note that if $\bar e$ is very small or the time step is very large, i.e., $\Delta t >> \frac{1}{\bar e L_N}$, adding and removing agents will dominate.
\end{corollary}

\begin{figure}
    \centering
    \includegraphics[width=0.55\linewidth]{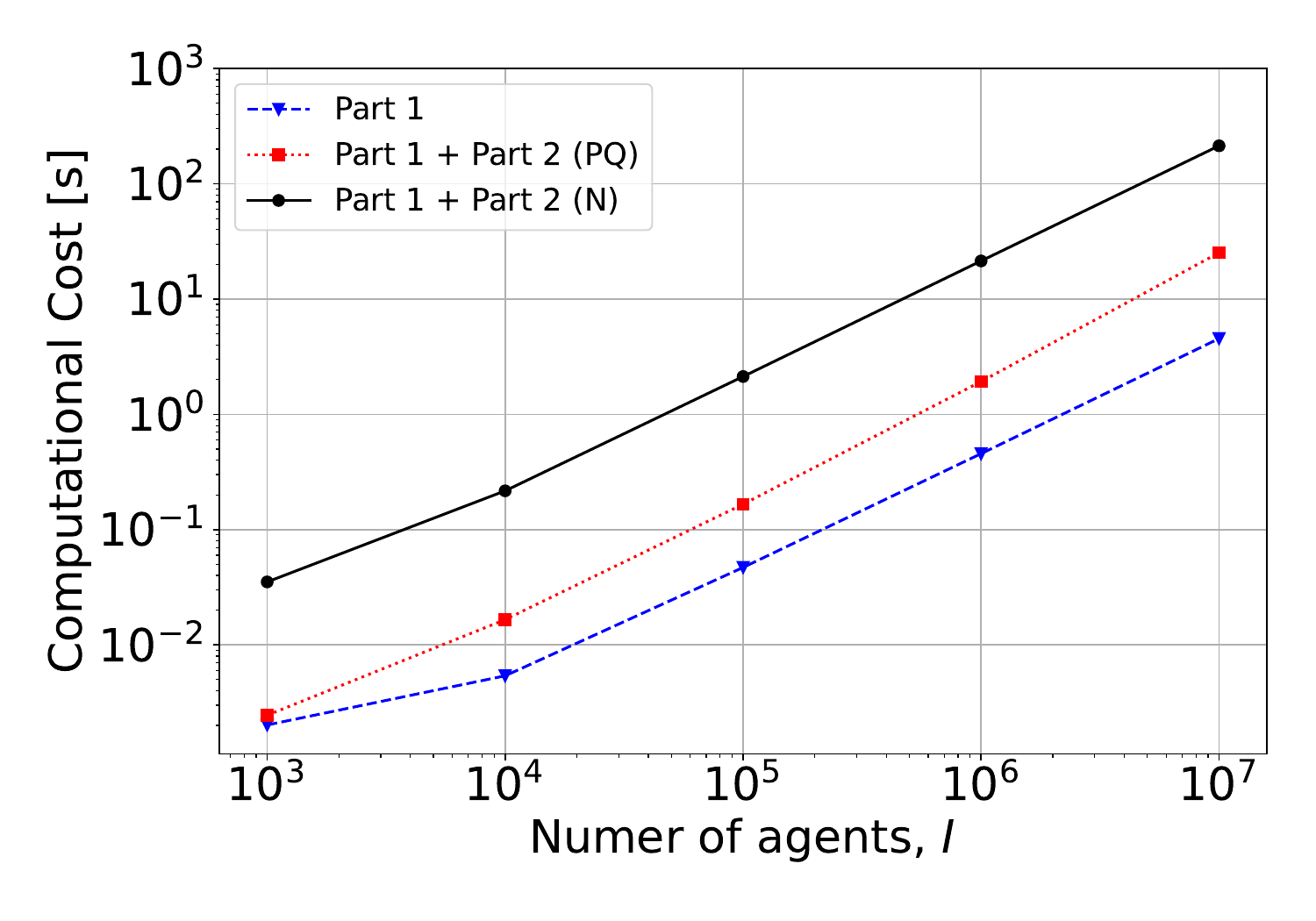}
    \caption{ \small Comparing Naive (N) and Priority Queue (PQ) formulations for 30 min of simulation, with $\Delta t =20$ s.}
    \label{fig:comparesorting}
\end{figure}

From Theorems \ref{theorem2} and \ref{theorem3}, the computational complexity of Algorithm \ref{Algorithm_heap} is significantly lower than a naive algorithm without exploiting the SCDFO principle of the bathtub model. The main difference is caused due to the method of calculating the completion rate fo trips. In the naive algorithm,  \refe{eq/g_assum} is used each time step to compare $E(t)$ times whether $x(t, i)<z(t)$. 
\reff{fig:comparesorting} presents a comparison of computational costs. For example, for 10 million agents, Part 1 is less than 5 s, Algorithm \ref{Algorithm_heap} has a cost of 20 s, and the Naive formulation is more than 200 s.

As discussed in Section \ref{sec:speed_variation}, when using a fixed time-step algorithm, one wants to ensure that the $\Delta t$ is small enough to eliminate numerical errors and captures adequately the changes in density in the system. Note that there is no requirement on the time step $\Delta t$ for the AB$^2$M to be well defined. $\Delta t$ should be only small enough to eliminate numerical errors. From Theorem \ref{theorem3}, it is clear that choosing a very small $\Delta t$ can increase the computational complexity of  Part 2 of the algorithm.
A first rough estimate for a time step can be obtained from $\Delta t \leq \frac{t_f}{I}$, which aims to capture on average the start of an agent. From  \refe{eq:e_average}, we have
\begin{equation}\label{eq:dt:Ln}
    \Delta t \leq  \frac{1}{\bar e L_N},
\end{equation}
Thus, scenarios with (or periods of) higher average inflow will require smaller $\Delta t$. In fact, for a fixed simulation period $t_f$, it is reasonable to decrease hyperbolically $\Delta t$ with an increasing number of agents, $I$, to maintain the same accuracy. However, from Section \ref{sec:speed_variation}, larger networks can allow more agents to enter or leave the system by maintaining the same accuracy (when using $\Delta v$ as a proxy). 
In all, there is a non-trivial relation on how to set up $\Delta t$ given the demand and supply characteristics. This complex relation goes beyond a trade-off between accuracy and computational cost when considering multiple parameters, such as the network length. Note that a reasonable lowest threshold of $\Delta t$ would be in the order of seconds. The exact selection of $\Delta t$ should be established after performing a convergence analysis for each scenario.

The flow-based scaling can further is expected to further reduce the computational cost. 
From Theorem \ref{theorem2}, the algorithm for a single simulation run in the scaled system has the complexity $\mathcal{O}(\frac{ I \log (r I)}{ \bar e \Delta t L_N})$\footnote{Note that some of the scaling effects are canceled out $\mathcal{O}(\frac{ rI \log (r I)}{ \bar e  r L_N \Delta t})$.}. 
Moreover, scaling the system can have implications for the time step that should be chosen. In particular, downscaling the system allows one to have a larger time step and still capture accurately the inflow of each individual vehicle following \refe{eq:dt:Ln}. 
Thus, considering $\Delta t_r = \frac{1}{\bar e L_N r}$, can further reduce the computational complexity as $\mathcal{O}( rI \log (r I))$.
Thus, a downscaled network $r<1$ can have a significant computational benefit.

\section{Conclusion}\label{sec:conclusion_ABM1}


The use of bathtub models, a.k.a. reservoir models, is becoming increasingly popular among researchers. The main difference from the traditional transportation models is that the bathtub model does not require setting up the physical network or tracking the agents' location within the network. Instead, the bathtub models capture the network flow dynamics in a relative space with respect to the trips' destinations. These bathtub models simplify the calibration and improve the computational efficiency through the elimination of vehicle position tracking and utilizing a global speed assumption. Additionally, bathtub models preserves privacy by avoiding the collection of personally identifiable location information.
In this study, we presented the agent-based bathtub model (AB$^2$M), which can be understood as a microscopic model that tracks the initiation, progression, and completion of individual trips in Lagrangian coordinates on relative space. This discrete version of the bathtub model has been overlooked in the literature because it is claimed to have a large computational cost, similar to traditional agent-based models in the absolute space \cite{KAGHO2020Agent}. 
However, using an agent-based formulation to describe the bathtub model traffic dynamics has the advantages that agents heterogeneity can be easily incorporated, and one could study the travel time reliability and other variables based on the higher-order moments obtained from Monte Carlo simulations.

We made two contributions to enhance the efficiency of the AB$^2$M:
First, in Section \ref{sec:efficiency} we leverage the ``shorter-(characteristic)-distance-first-out" (SCDFO) principle \citep{jin2020generalized} to introduce $\Theta(t,n)$ as a sorted collection of active trips based on their characteristic trip distance. This variable plays a vital role to propose an efficient algorithm that uses $\Theta(t,n)$ as a priority queue with binary trees.
Secondly, in Section \ref{sec:scalability} we examine the scalability of bathtub models and propose two downsizing approaches: the flow-based downscaling, which reduces the number of agents without introducing biases into the trip flow dynamics, and the distance-based downscaling, which is proven to be equivalent to the original system only in steady states. However, the existence and reachability of such steady states is out of the scope of this paper.  Numerical results showcase that if the scaling ratio in the distance-based downscaling is very low or very large the dynamics will differ significantly.
Then, a systematic discussion on the computation complexity of the algorithms presented in Section \ref{sec:complexity}.
Finally, through numerical simulations, we discussed the differences between the AB$^2$M and two of the most established continuum bathtub models, i.e., the Vickrey's bathtub model (VBM) \citep{Vickrey2020congestion}, sometimes referred to as accumulation-based model \citep{MARIOTTE2017} or PL model \citep{SIRMATEL2021Modeling}, and the Generalized bathtub model \citep{jin2020generalized}, which is equivalent to the Trip-based model \citep{MARIOTTE2017} for time-independent trip distance distributions. 
Through numerical examples, we also study the higher-order moments, such as trip travel time distribution (TTTD), for three different common assumptions on the trip distance distribution, i.e., the NE distribution \citep{Vickrey2020congestion}, the constant trip distance \citep{Arnott1993, ARNOTT2018}, the log-normal distribution \citep{Martinez2021On}. We showed that for the same average trip distance, the NE distribution leads to less severe congestion and shorter travel times than systems with constant trip distance or with a log-normal distribution. This highlights the importance of not relying on VBM to develop management strategies at the network level since they could underestimate the travel time of users.
Further, the numerical simulations also shed some light on how to select $\Delta t$ and $r$ for a down-scaled city.


An AB$^2$M has several advantages over traditional ABM in the absolute space: First, it is computationally more efficient, because one does not need to track the position of vehicle-trips in the real network, which ensures the preservation of personally identifiable location information, addressing privacy concerns associated with individual data collection in ABM in absolute space. Secondly, the speed of all vehicles is the same (global speed), determined by a single equation based on the NFD assumption. In contrast, in the absolute space, the speed is local, usually determined at the link level.  
Thirdly, the flow-based downscaling employed in AB$^2$M introduces no biases and exhibits no physical limitations on the extent of downsizing the system.

In summary, the main contributions of this paper are threefold:
\begin{itemize}
    \item[1.] A methodological/modeling contribution, i.e., the definition of $\Theta(t,n)$ as a priority queue (through binary trees) of active agents sorted by characteristic trip distance, build on the SCDFO principle. 
    \item[2.] Although the downscaling of agents is a natural idea and has been implemented in the past, this paper thoroughly discuses two scaling implementations as well as their impacts both analytically and numerically. We show that the flow-based downscaling of  AB$^2$M allows to reduce the computational cost without introducing biases. 
    \item[3.]  Rigorously established the upper bound of the computational complexity of the two fixed time step formulations of the AB$^2$M, and the benefits of donwscaling.
\end{itemize}

This study is the first step towards a better understanding of agent-based modeling in the relative space, the impacts of downscaling and time-step discretization on the computational efficiency and numerical errors. From the insights gained in this paper, an event-based formulation of the AB$^2$M, as the ones proposed by \cite{MARIOTTE2017, Lamotte2018}, may consider longer event-periods than the start or end time of each trip and still eliminate the numerical errors. In the future, we are interested in a detailed comparison of accuracy and computational complexity between a discrete time step and an event-based formulation of the AB$^2$M.

The proposed AB$^2$M considered a single bathtub and single transportation mode (and thus a single speed for the whole set of trips at a given time). In the future, the AB$^2$M should also be extended to study  multi-region transportation systems with multiple connected bathtubs. To model several connected bathtubs, e.g., modeling the downtown and periphery of a city with different bathtubs, the extension of the AB$^2$M should define how to handle the trip moving between the connected bathtubs.

Moreover, due to its computational efficiency, the AB$^2$M has the potential to be extended to model multi-modal transportation systems and incorporate the decision making of independent agents based on behavioral rules. Due to the nature of agent-based modeling, AB$^2$M can be extended to systems where the person trips and vehicle-trips are not the same.
While the number of vehicles in the system determines the speed (i.e., rate of trip progression), the passenger trips could be different from the vehicle-trips. For example, considering transit services where vehicles generally have high occupancy and the trip of users does also involve access and waiting time. In this case, the agents' trips can be modeled as a chain of trips, i.e., the trip consists of several stages or ``states''. These agents' states can be defined as a set or subset of attributes (e.g., the agent can be at her origin/destination, walking to the bus station, waiting for a bus, or traveling inside the bus). Each state of its trip has a different speed progression. The possible agent states should be defined depending on the transportation system to be modeled. An example of these states is presented in the compartmental model, where trips are planned, traveling, or completed for privately owned vehicle-trips or can be planned, waiting, traveling, and completed for users of a shared mobility system \citep{Jin2021Compartmental}. The trip chains can be modeled as different compartments or bathtubs. 

Note that even if the downscaling (or up-scaling) in the relative space framework has been shown to be mathematically rigorous with a single-mode and single bathtub, the introduction of other modes and multiple bathtubs may require more discussions and extensions of the scaling properties. How this multi-modal system should be downscaled has not been systematically studied yet; neither in the absolute  \citep{BENDOR2021Population} nor relative spaces. In fact, multi-modal ABM present a significant problem for downscaling in the absolute space. Future studies on downscaling should try to address these issues.

\small 

\subsection*{Acknowledgments}
The authors would like to thank the support of the California Statewide Transportation Research Program (SB1), 2021-2022, and the SCC-IRG Track 1 from NSF-SCC CMMI 2125560.

\bibliography{thesis}
\end{document}